\documentclass[a4paper,aps,prx,showpacs,twocolumn,superscriptaddress]{revtex4-2}

\usepackage[utf8]{inputenc}  
\usepackage[T1]{fontenc}     
\usepackage[british]{babel}  
\usepackage{lmodern}  

\usepackage[scaled=1.03]{inconsolata} 
\usepackage[usenames,dvipsnames]{color} 
\usepackage[colorlinks,citecolor=blue,linkcolor=magenta,urlcolor=blue]{hyperref}  
\usepackage{graphicx} 
\usepackage{tikz}
\usepackage[babel]{microtype}  
\usepackage{amsmath,amssymb,amsthm,bm,mathtools,amsfonts,mathrsfs,bbm,dsfont} 
\usepackage{xspace}  
\usepackage{pgf}
\usepackage{newtxtext}
\usepackage{newtxmath}
\usepackage[cal=cm]{mathalfa}

\usepackage{multirow}
\usepackage{verbatim}
\usepackage{enumerate}
\usepackage{tcolorbox}
\usepackage[normalem]{ulem}

\usepackage{physics}
\usepackage[mathscr]{eucal}

\usepackage{xcolor}
\definecolor{memory}{HTML}{d948bc}

\makeatletter
\let\comment\relax

\makeatother
\providecommand{\comment}[1]{}

\usepackage{tikz,quantikz}
\usepackage{tikz}
\usetikzlibrary{patterns}
\usetikzlibrary{patterns.meta}

\tikzdeclarepattern{
  name=hatch,
  parameters={\hatchsize,\hatchangle,\hatchlinewidth},
  bounding box={(-.1pt,-.1pt) and (\hatchsize+.1pt,\hatchsize+.1pt)},
  tile size={(\hatchsize,\hatchsize)},
  tile transformation={rotate=\hatchangle},
  defaults={
    hatch size/.store in=\hatchsize,hatch size=5pt,
    hatch angle/.store in=\hatchangle,hatch angle=0,
    hatch linewidth/.store in=\hatchlinewidth,hatch linewidth=.4pt,
  },
  code={
      \draw[line width=\hatchlinewidth] (0,0) -- (\hatchsize,\hatchsize);
  }
}

\tikzdeclarepattern{
  name=mylines,
  parameters={
      \pgfkeysvalueof{/pgf/pattern keys/size},
      \pgfkeysvalueof{/pgf/pattern keys/angle},
      \pgfkeysvalueof{/pgf/pattern keys/line width},
  },
  bounding box={
    (0,-0.5*\pgfkeysvalueof{/pgf/pattern keys/line width}) and
    (\pgfkeysvalueof{/pgf/pattern keys/size},
0.5*\pgfkeysvalueof{/pgf/pattern keys/line width})},
  tile size={(\pgfkeysvalueof{/pgf/pattern keys/size},
\pgfkeysvalueof{/pgf/pattern keys/size})},
  tile transformation={rotate=\pgfkeysvalueof{/pgf/pattern keys/angle}},
  defaults={
    size/.initial=5pt,
    angle/.initial=45,
    line width/.initial=.4pt,
  },
  code={
      \draw [line width=\pgfkeysvalueof{/pgf/pattern keys/line width}]
        (0,0) -- (\pgfkeysvalueof{/pgf/pattern keys/size},0);
  },
}

\usetikzlibrary{arrows.meta, shadings}

\usepackage{pgfplots}
\usepackage{tikz-3dplot}
\tdplotsetmaincoords{60}{115}
\tdplotsetrotatedcoords{25}{-25}{0}
\pgfplotsset{compat=newest}
\usetikzlibrary{arrows.meta, arrows, positioning}

\pgfplotsset{compat = newest}

\newcommand{\blochdyn}[8]{

	\shade[ball color = lightgray,opacity = 0.2] (0,0,0) circle (1cm);
	\tdplotsetrotatedcoords{#6}{#7}{#8};
	
	\shade[tdplot_rotated_coords,right color=orange,middle color=red,left color=blue,opacity=0.3,shading angle=-110] (#1,#2,#3) circle (#4 and  #5);
	\shade[tdplot_rotated_coords,ball color=red,opacity=0.5] (#1,#2,#3) circle (#4 and #5);
	
	\tdplotsetrotatedcoords{0}{0}{0};
	\draw[dashed,tdplot_rotated_coords,gray] (0,0,0) circle (1);
	
	\tdplotsetrotatedcoords{90}{90}{90};
	\draw[dashed,tdplot_rotated_coords,gray] (1,0,0) arc (0:360:1);
	
	\tdplotsetrotatedcoords{0}{90}{90};
	\draw[dashed,tdplot_rotated_coords,gray] (1,0,0) arc (0:360:1);
	
	\draw[-{Latex[length=2mm]}] (0,0,0) -- (1,0,0);
	\draw[-{Latex[length=2mm]}] (0,0,0) -- (0,1,0);
	\draw[-{Latex[length=2mm]}] (0,0,0) -- (0,0,1);
	\draw[dashed, gray] (0,0,0) -- (-1,0,0);
	\draw[dashed, gray] (0,0,0) -- (0,-1,0);
	\draw (1.2,-.1, 0.1) node[anchor=north west] {$x$};
	\draw (0, 0.95, 0.2) node[anchor=north west] {$y$};
	\draw (0, 0, 1) node[anchor=north west] {$z$};
    
    }

\definecolor{mathematicablue}{rgb}{0.87,0.94,1}
\definecolor{mathematicadarkblue}{rgb}{0.368417, 0.506779, 0.709798}
\definecolor{mathematicaorange}{rgb}{1,0.9,0.8}
\definecolor{mathematicadarkorange}{rgb}{1,0.5,0}
\definecolor{darkyellow}{rgb}{1,0.8,0}
\definecolor{darkred}{rgb}{0.85,0.04,0}
\definecolor{niceblue}{rgb}{0.165,0.831,1.}
\definecolor{nicegreen}{rgb}{0.494,0.835,.082}

\newcommand{\id}{\ensuremath{\vvmathbb{1}}}

\newcommand{\kb}[2]{|#1\rangle\langle#2|} 

\newcommand{\Ecal}{\mathcal{E}}

\newcommand{\Dcal}{\mathcal{D}}
\newcommand{\Xcal}{\mathcal{X}}
\newcommand{\Ycal}{\mathcal{Y}}
\newcommand{\Asf}{\mathsf{A}}
\newcommand{\Bsf}{\mathsf{B}}
\newcommand{\Csf}{\mathsf{C}}
\newcommand{\Dsf}{\mathsf{D}}

\newcommand{\Sys}{\mathcal{P}} 
\newcommand{\Anc}{\mathcal{A}} 
\newcommand{\Mem}{\mathcal{M}} 

\newcommand{\Dyn}{\mathcal{D}}

\newcommand{\cpt}{\Ecal}

\newcommand{\phaseGate}{S}

\newcommand{\mP}{m}
\newcommand{\mM}{M}
\newcommand{\xP}{x}
\newcommand{\xM}{X}
\newcommand{\aP}{a}
\newcommand{\aM}{A}
\newcommand{\qP}{q}
\newcommand{\qM}{Q}
\newcommand{\pP}{p}
\newcommand{\pM}{P}
\newcommand{\omegaP}{\omega}

\newcommand{\cov}{\boldsymbol{\sigma}}
\newcommand{\Sigmacov}{\boldsymbol{\Sigma}}
\newcommand{\alphacov}{\boldsymbol{\alpha}}
\newcommand{\betacov}{\boldsymbol{\beta}}
\newcommand{\gammacov}{\boldsymbol{\gamma}}

\makeatletter
\def\maketitle{
\@author@finish
\title@column\titleblock@produce
\suppressfloats[t]}
\makeatother

\usepackage{chngcntr}

\pdfpageattr{/Group << /S /Transparency /I true /CS /DeviceRGB >>}

\begin{document}
\title{One-sided Witnesses for the Quantumness of Gravitational Dynamics}
\author{Konstantin Beyer}
\affiliation{Department of Physics, Stevens Institute of Technology, Hoboken, New Jersey 07030, USA}
\author{M. S. Kim}
\affiliation{Blackett Laboratory, Imperial College London, London, SW7 2AZ, United Kingdom}
\author{Igor Pikovski}
\affiliation{Department of Physics, Stevens Institute of Technology, Hoboken, New Jersey 07030, USA}
\affiliation{Department of Physics, Stockholm University, Stockholm, Sweden}

\date{\today}

\begin{abstract}
Quantum information concepts and quantum technologies have opened the prospect of probing the quantum nature of gravity in table-top experiments. Most proposals rely on detecting gravitationally mediated entanglement, which requires measurements on both interacting systems. Here we propose an alternative test based on witnessing quantum memory effects in the dynamics of a single system. We show how the quantum memory concept allows the non-classicality of the interaction between two systems to be certified through local measurements on only one of them. We then develop a novel quantum memory witness, making the approach applicable to gravitationally coupled harmonic oscillators. This witness requires no assumptions about the dynamics, such as Gaussianity. Our results show that the quantumness of gravitational dynamics can be certified with access to only one subsystem, opening the door to asymmetric implementations that can offer conceptual and experimental advantages to probing the quantum nature of the gravitational interaction.
\end{abstract}

\maketitle

\section{Introduction}
The idea of testing the quantumness of gravity in table-top experiments has attracted a lot of attention in recent years \cite{marshall2003towards,pikovski2012probing,bekenstein2012tabletop,yang2013macroscopic,kafri2014classical,belenchia2016testing,pfister2016universal,krisnanda2017revealing,carney2019tabletop,bose_spin_2017,marletto_gravitationally_2017,chevalier_witnessing_2020,guff2022optimal,krisnanda_observable_2020,carneyUsingAtomInterferometer2021,matsumura_leggett-garg_2022,carlesso2019testing,anastopoulos2020quantum,danielson2022gravitationally,pedernales2022enhancing,christodoulou2023locally,kaku2023enhancement,hanif2024testing,bose2025massive}.
The Colella-Overhauser-Werner experiment in 1975~\cite{colella_observation_1975} and the neutron bouncing experiment in 2002~\cite{nesvizhevsky_quantum_2002} demonstrated that non-relativistic quantum particles in an external gravitational field are well described by the Schr\"{o}dinger equation with a Newtonian potential. 
However, performing an experiment which shows that quantum particles also \emph{source} gravity in the very same way is much harder, since it basically means to measure the gravitational field of a large mass in quantum superposition. 
While there is a deep link between quantum probes and quantum sources of gravity that allows for new indirect tests \cite{fragkos2025probing}, a direct experimental proof of gravitating quantum sources would be a major milestone to verify the expected quantum behavior, and rule out possible alternatives like semiclassical gravity~\cite{bahrami2014schrodinger,anastopoulos2015probing,oppenheim_postquantum_2023, angeliProbingQuantumNature2025}. Other than the direct detection of gravitons \cite{tobar2024detecting,shenderov2024stimulated},  probing quantum sources of gravity is thus one of the promising directions for future table-top experiments exploring the quantum nature of gravity.

\begin{figure}
    \centering
    \begin{tikzpicture}

    \node at (-2.2,.5) {(a)};
    \node at (2.5,.5) {(b)};

\draw[|-, line width = .5 mm] (1.2,.6) -- (1.2,-.7);
  \draw[fill = memory, line width = .5mm] (1.2,-.7) circle (.7cm) node {\(M\)};

\draw[|-, line width = .5 mm] (-1,.6) -- (-1,-.7);
  \draw[fill = darkyellow, line width = .5mm] (-1,-.7) circle (.3cm) node {\(m\)};

\draw[<->, line width = .3mm] (-.7,-.7) -- (.5,-.7) node[midway, below]{\(D\)};

\draw[<->, line width = .3mm] (-1,.25) -- (1.2,.25) node[midway, below]{\(d\)};

\draw[|-|, line width = .3mm] (-1.3,-1.2) -- (-.7,-1.2) node[midway, below]{\(2 r\)};

  \draw[|-, line width = .5 mm] (4.5,.6) -- (4.5,-.3);
  \draw[fill = memory, line width = .5mm] (4.5,-.3) circle (.4cm) node {\(M\)};
  \draw[fill = darkyellow, line width = .5mm] (3.5,-.7) circle (.2cm) node {\(m\)};
  \node at (3.5,-.2) {\(\ket{l}\)};
  \draw[dashed, fill = darkyellow, line width = .5mm] (5.5,-.7) circle (.2cm);
  \node at (5.5,-.2) {\(\ket{r}\)};
  \draw[|-|, line width = .3mm] (3.5,-1.2) -- (5.5,-1.2) node[midway, below]{\(d\)};

\end{tikzpicture}
    \caption{Two prototypical setups to probe a gravitational channel. (a) Two harmonic oscillators with masses \(M\) and \(m\). (b) A spatial qubit of mass \(m\) interacts with a harmonic oscillator of mass \(M\).
    }
    \label{fig:setups}
\end{figure}

\begin{figure}
\centering
\includegraphics[width = \linewidth]{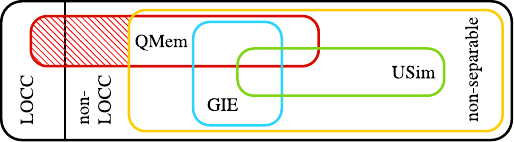}








\caption{Schematic diagram of the relation of different classes of CPT maps and potential witnesses of quantum dynamics. 
Although non-LOCC is often considered as a definition for quantumness of the joint dynamics, tractable witnesses actually only work on the subset of non-separable dynamics, i.e., those that cannot be decomposed into Kraus operators of product form. {GIE} (blue) denotes the witnesses based on entanglement detection. USim  (green) refers to witnesses based on the simulability of the expected unitary dynamics by LOCC maps~\cite{lamiTestingQuantumnessGravity2024a,toccaceloBenchmarksQuantumCommunication2025a}. Our quantum memory witness (QMem) also detects a subset of the non-separable maps (red), while requiring operations and measurements only on the probe system \(\Sys\). In principle, this witness could also detect cases  where the gravitational field itself provides the quantum memory. In this situation, the quantumness of gravity would be verified regardless of whether the joint map of \(\Sys\) and \(\Mem\) is LOCC (see App.~\ref{app:LOCC} for details).      
However, this case (hatched region) is not expected in a table-top setup. 
The expected unitary dynamics predicted by the Schrödinger equation lies in the intersection of all witnesses.    }
\label{fig:classes}
\end{figure}

In order to draw conclusions about gravity acting between quantum source masses, it is necessary to measure observables that witness the quantized gravitational interaction between such systems. A typical setup consists of gravitationally interacting mechanical oscillators or masses in spatial superposition (see Fig.~\ref{fig:setups})~\cite{bose_spin_2017,marletto_gravitationally_2017,chevalier_witnessing_2020,guff2022optimal,krisnanda_observable_2020,carneyUsingAtomInterferometer2021,matsumura_leggett-garg_2022}.
The key signature often considered is the build-up of entanglement through the gravitational interaction, or \emph{gravitationally induced entanglement} (GIE)~\cite{bose_spin_2017,marletto_gravitationally_2017}. 
The reasoning follows from the fact that a classical mediator picture, described by local operations and classical communication (LOCC), cannot generate any entanglement \cite{horodecki2009quantum,chitambar2014everything}. Variations of this reasoning include tests of the simulability of the expected unitary evolution~\cite{lamiTestingQuantumnessGravity2024a,toccaceloBenchmarksQuantumCommunication2025a}, also providing witnesses of non-LOCC dynamics for gravity (see Fig.~\ref{fig:classes}). While the verification of a non-LOCC channel is not necessarily a proof of quantized underlying degrees of freedom of gravity \cite{hall2018two,anastopoulos2018comment,anastopoulos2020quantum,fragkos2022inference}, it nevertheless shows that the mutual gravitational interaction must be of quantum nature when source masses are quantized. It essentially verifies that the Newtonian potential is sourced in a superposition 
according to $\hat{\Phi}(\vec{x})=- G m\int \! d^3\!y\, \hat{\psi}^{\dagger}({\vec{y}}) \hat{\psi}({\vec{y}})/|\vec{x}-\vec{y}| $, where $\hat{\psi}^{\dagger}(\vec{x}) \ket{0} = \ket{\vec{x}}$ is an operator in the second quantized picture that corresponds to mass $m$ at position $\vec{x}$ \cite{fragkos2022inference}. The LOCC test is thus a test of the quantum nature of the source and that the mutual interaction proceeds according to unitary quantum dynamics \cite{krisnanda2017revealing,fragkos2022inference,carney2022newton}. (There is an additional interpretation that the experiment probes the quantum nature of virtual particles \cite{bose_spin_2017,marletto_gravitationally_2017,danielson2022gravitationally,christodoulou2023locally}. However, this perspective relies on additional assumptions about the ontological mechanism of entanglement generation, which is neither experimentally accessible nor uniquely determined even in relativity \cite{franson2011entanglement,anastopoulos2018comment,anastopoulos2020quantum,fragkos2022inference,martin-martinez_what_2023,aziz_classical_2025}.)

Established proposals of testing for non-LOCC dynamics as a signature of the quantum nature of the interaction {have the shortcoming of being} 
inherently based on measures of correlation between the two gravitationally interacting masses.
State preparation and/or quantum measurements on both subsystems are mandatory in order to verify the entanglement or non-LOCC witnesses.
On the other hand, while local signatures like the decoherence of the subsystems~\cite{carneyUsingAtomInterferometer2021}, the violation of suitable Leggett-Garg inequalities~\cite{matsumura_leggett-garg_2022}, {or tests based on conserved quantities~\cite{di_pietra_temporal_2023,pietra_temporal_2025}} can be demonstrated by measuring only one of the gravitationally interacting subsystems, such one-sided approaches are not sufficient for proving the quantum nature of the underlying dynamics, as they {either rely on additional strong assumptions or} can equally be explained by classical models of the dynamics~\cite{hosten_constraints_2022,maLimitsInferenceGravitational2022,streltsov_significance_2022,carney_erratum_2022}. Thus, these signatures are necessary if gravity is quantum but they are not sufficient, which is most intuitively evident from the fact that  they become stronger under classical thermal noise~\cite{carneyUsingAtomInterferometer2021,matsumura_leggett-garg_2022}. 

In this paper, we develop an alternative test of a quantum signature of gravity that is both one-sided \emph{and} sufficient. Loosely speaking, the signature tests if gravity can write and read a quantum memory, which cannot be achieved by an underlying classical dynamics.
More concretely, we focus on the reduced dynamics of one of the two gravitationally interacting systems, which we call the \emph{probe} \(\Sys\). The other system serves as a quantum \emph{memory} \(\Mem\).
Based on concepts of Refs.~\cite{backerLocalDisclosureQuantum2024,backerEntropicWitnessQuantum2025a,yu_quantum_2025} that were developed for quantum signatures in non-Markovian open systems, we show how suitable sets of measurements on the probe system at different points in time can witness the coherent transfer of quantum information through gravity between the two subsystems at intermediate times. 
Unlike previously proposed one-sided signatures, our witness rules out classical dynamics. 
A successful test shows that either the gravitational interaction is non-LOCC or that the gravitational field itself provides quantum memory effects (see Fig.~\ref{fig:classes}).
The theoretical concepts developed here open up new routes for experimental approaches that are inherently asymmetric and probe a genuine quantum feature beyond entanglement generation---which, as we show, may offer significant 
experimental advantages in certain parameter regimes. 

\section{Gravitationally coupled quantum memory}
\begin{figure}[t]
    \centering
    \includegraphics[width=.6\linewidth]{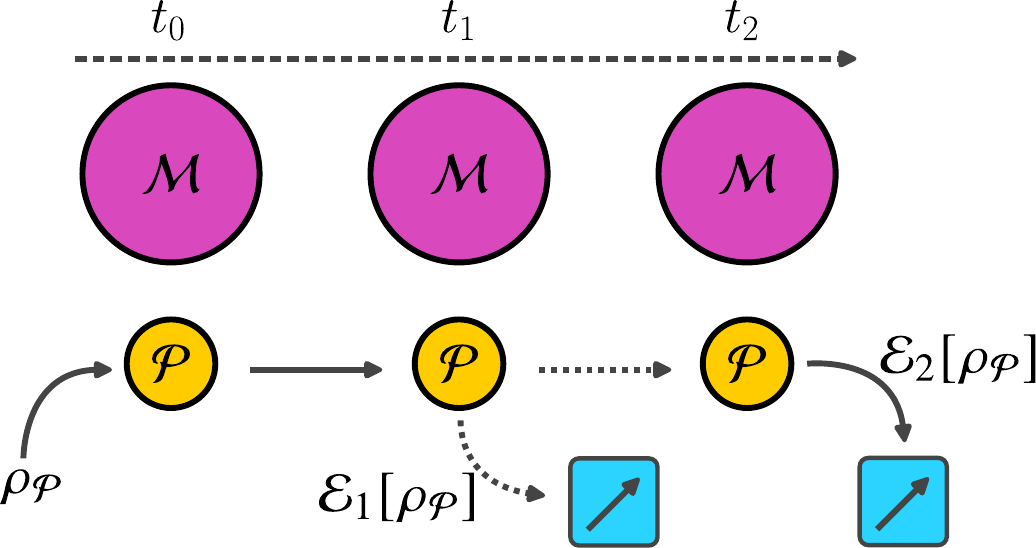}
    \caption{The basic scheme to verify non-LOCC gravitational dynamics with a single probe system. The probe \(\Sys\) interacts gravitationally with another quantum system \(\Mem\). The aim is to verify by means of measurements on \(\Sys\) alone that gravity can coherently transfer quantum information between the two systems. To this end the experimenter prepares various initial probe states \(\rho_\Sys\) at time \(t_0\) and measures the resulting state at either time \(t_1\) or \(t_2\). The probe state at the time of the measurements is given by the local maps \(\cpt_1\) and \(\cpt_2\), respectively. By measuring suitable observables, the experimenter can verify that the measurement outcomes at time \(t_2\) can only be explained if quantum information has been stored in \(\Mem\) at the intermediate time \(t_1\), proving that gravity must have coherently transferred this information. Importantly, the quantum witness is truly one-sided as \(\Mem\) is never measured.}
    \label{fig:intro}
\end{figure}

The basic scheme is shown in Fig.~\ref{fig:intro}. We consider two gravitationally interacting quantum systems \(\Sys\) (probe) and \(\Mem\) (memory). All other forces are assumed to be negligible.
The probe is well controlled. In particular, we assume that the preparation of different initial states and the measurement of various observables can be implemented on \(\Sys\).
By contrast, no detailed knowledge about \(\Mem\) is required: the experimenter can, in principle, remain ignorant about the Hilbert space of \(\Mem\) and no measurements are performed on this subsystem.

The two subsystems start in a product state \(\rho=\rho_\Sys \otimes \rho_\Mem\). While the experimenter can choose \(\rho_\Sys\) at will, the initial state of the memory \(\rho_\Mem\) may in principle be hidden. We only assume that it is statistically independent of the state prepared in \(\Sys\). 
The gravitational interaction between \(\Sys\) and \(\Mem\) generates a dynamics \(\Xi_t\), i.e., a family of maps that propagates the joint state \(\rho\) at time \(t_0\) to a state \(\rho_t\) at \(t>t_0\).
In order to obtain physically valid output states, the maps \(\Xi_t\) have to be completely positive and trace-preserving (CPT).
This joint dynamics \(\Xi_t\) induces a local CPT dynamics \(\cpt_t\) on \(\Sys\) alone, given by the trace over the part \(\Mem\), i.e.,
\begin{align}
    \cpt_{t}[\rho_\Sys] = \tr_\Mem\left[\Xi_{t}(\rho_\Sys \otimes \rho_\Mem) \right].
\end{align}
While \(\Xi_t\) cannot be measured directly due to the lack of detailed information about \(\Mem\), the local dynamics \(\cpt_t\) can be experimentally determined in the setting considered here by channel tomography on the probe \(\Sys\).

It was shown in Ref.~\cite{backerLocalDisclosureQuantum2024} that the \emph{local} dynamics \(\cpt_t\) in \(\Sys\) bears valuable information about the nature of the \emph{global} dynamics \(\Xi_t\). 
Let us assume that the experimenter takes a snapshot of the time evolution by performing process tomography for two different times \(t_{1}\) and \(t_2\), obtaining two local maps \(\cpt_1\) and \(\cpt_2\) on \(\Sys\).     
The dynamics \(\Dcal = (\cpt_1,\cpt_2)\) is \emph{realizable with classical memory} if it can be decomposed as~\cite{backerLocalDisclosureQuantum2024}
\begin{align}
    \label{eq:classical-memory}
    \cpt_1[\rho_\Sys] = \sum_i K_i \rho_\Sys K_i^\dagger, &&
    \cpt_2[\rho_\Sys] = \sum_i \Phi_i[K_i \rho_\Sys K_i^\dagger],
\end{align}
where \( \sum_i K_i^\dagger K_i =\id\), and the \(\Phi_i\) are CPT maps.
The Kraus operators \(K_i\) can be thought of as a measurement of the system with outcome \(i\) which on average realizes the map \(\cpt_1\) at \(t_1\). In a second step, CPT maps \(\Phi_i\) conditioned on the previous outcome \(i\) are applied, which on average realize the map \(\cpt_2\) at time \(t_2\).
The crucial point is that the outcome label \(i\), which is needed to condition the map \(\Phi_i\) in step two, is classical data storable in a classical memory.
Thus, dynamics \(\Dcal\) that can be decomposed as in Eq.~\eqref{eq:classical-memory} are, in principle, realizable without employing quantum memory effects, despite possibly being non-Markovian. 
More importantly, if the dynamics of the probe \(\Sys\) \emph{cannot} be written in this way, there must be a quantum memory system which stores quantum information at time \(t_1\) that is retrieved later to realize the map \(\cpt_2\) at time \(t_2\).

Eq.~\eqref{eq:classical-memory} is crucial for what follows. Ruling out such a classical decomposition for the probe dynamics, resulting from the gravitationally coupled scenario considered here, would demonstrate a strong quantum signature of gravity. It would show that either the gravitational interaction is able to coherently transfer quantum information into the quantum memory \(\Mem\) and read it out at a later time, or that gravity itself provides a quantum memory for the dynamics (although the second option seems unlikely in the table-top regime considered here). 
We note that the observation of quantum memory effects in the local dynamics would also directly prove the non-LOCC behavior of the global gravitational dynamics (see App.~\ref{app:LOCC}).

Finding a classical memory decomposition of the form in Eq.~\eqref{eq:classical-memory} for a given dynamics \(\Dyn=(\cpt_1,\cpt_2)\) is a computationally hard problem.
However, its non-existence and therefore the quantumness of the gravitational interaction can be witnessed by several tractable sufficient criteria~\cite{backerLocalDisclosureQuantum2024,backerEntropicWitnessQuantum2025a,yu_quantum_2025,backerVerifyingQuantumMemory2026,backerRevealingQuantumNature2026}. 
The choice of a suitable quantum witness strongly depends on the Hilbert space of the probe \(\Sys\) and the dynamics to be tested (similar to an entanglement witness). 
In the following we present two exemplary cases that use an oscillator and a qubit probe, respectively.

\section{Two gravitationally coupled oscillators}

We assume the probe \(\Sys\) and the memory \(\Mem\) to be two quantum harmonic oscillators with spherical masses \(\mP\) and \(\mM\), displacements \(\xP\) and \(\xM\), respectively, with resonant frequencies \(\omegaP\).
Let \(d\) be the distance between the rest center-of-mass positions of the two oscillators [see Fig.~\ref{fig:setups}\;(a)]. The gravitational interaction energy is then given by \(H_g = -G \mP \mM/\abs{d + (\xP - \xM)}\).
Expanding the potential, neglecting global phases, and absorbing constant frequency offsets into \(\omega\), the dominating quantum coupling term between the two oscillators reads
\(H_g = 2 G \mP \mM \xP \xM  /d^3\).
Writing \(\xP = \sqrt{\hbar/(2m\omega)}(\aP^\dagger + \aP)\),  \(\xM = \sqrt{\hbar/(2 \mM \omega)}(\aM^\dagger + \aM)\) in terms of bosonic operators \(a, A\), applying the rotating-wave approximation, and going into a rotating frame with respect to the free oscillator frequency \(\omega\), we arrive at the Hamiltonian~\cite{lamiTestingQuantumnessGravity2024a,toccaceloBenchmarksQuantumCommunication2025a}
\begin{align}
\label{eq:osci-Hamiltonian}
    H = \hbar g \left(\aP^\dagger \aM + \aP \aM^\dagger\right), && g = \frac{G \sqrt{\mP \mM}}{d^3 \omega}.
\end{align}
Let us furthermore assume that both oscillators undergo a Markovian damping with rate \(\gamma\) into a bath with effective mean excitation \(\bar{N}\).
The expected dynamics can straightforwardly be solved in the symplectic formalism for Gaussian states~\cite{serafini_quantum_2023}. 
We define the dimensionless quadratures \(\qP = (\aP^\dagger + \aP)/\sqrt{2}\), \(\pP = i(\aP^\dagger - \aP)/\sqrt{2}\), \(\qM = (\aM^\dagger + \aM)/\sqrt{2}\), \(\pM = i(\aM^\dagger - \aM)/\sqrt{2}\), and the vector \(\vec\xi = (\qP,\pP, \qM, \pM)^\top \). 
A two-mode Gaussian state \(\rho_{\Sys\Mem}\) is characterized by the expectation values \(l_i = \langle \xi_i \rangle = \tr[\xi_i \rho]\), and the covariance matrix \(\cov_{ij} = \langle \xi_i\xi_j + \xi_j\xi_i \rangle/2 - \langle \xi_i \rangle \langle \xi_j \rangle\).

The Hamiltonian in Eq.~\eqref{eq:osci-Hamiltonian} can be written as 
\begin{align}
    H = \hbar \vec\xi^\top \Tilde{H} \vec\xi /2, && \Tilde{H} = g \sigma_x \otimes \id,
\end{align}
with the standard Pauli matrix \(\sigma_x\).
Under thermal damping with strength \(\gamma\) into a bath of average excitation \(\bar{N}\), the two-mode covariance matrix \(\cov_{\Sys\Mem}\) and displacement vector \(\vec l_{\Sys\Mem}\) evolve according to~\cite{serafini_quantum_2023,sandulescu_open_1987}
\begin{align}\label{eq:eom}
    \dot{\cov}_{\Sys\Mem} = K 
     \cov_{\Sys\Mem} + \cov_{\Sys\Mem} K^\top + D, && \dot{\vec{l}}_{\Sys\Mem} = K \vec{l}_{\Sys\Mem},
\end{align}
 where \(K = J \Tilde{H} - \gamma \id/2\), with \(J\) the symplectic matrix, and \(D = \gamma(\bar{N}+1/2)\id\).
The formal solution is given by~\cite{serafini_quantum_2023,sandulescu_open_1987}
\begin{align}
   \cov_{\Sys\Mem}(t) = X_t \cov_{\Sys\Mem}(0) X_t^\top + Y_t,\; \vec{l}_{\Sys\Mem}(t) = X_t \vec{l}_{\Sys\Mem}(0) 
\end{align} with 
\begin{align}
\label{eq:X-and-Y}
    X_t = \exp[K t], && Y_t = \int_{0}^t X_s D X_s^\top \mathrm{d}s,
\end{align} which can be computed in closed form for the case considered here.

We are interested in the reduced dynamics of the probe \(\Sys\),  \(\cov := \cov_\Sys\), \(\vec l := \vec l_\Sys\),  which is given by a transformation of the form
\begin{align}\label{eq:map1}
    \cpt_t: \quad \cov \mapsto W_t \cov W_t^\top + V_t, \quad \vec l \mapsto W_t \vec l + \vec \lambda_t. 
\end{align}
Writing \(X_t\) and \(Y_t\) in block form of \(2\times 2\) matrices  we have 
\begin{align}
    \cov_{\Sys\Mem}(t) = \begin{pmatrix}
        X_a & X_b \\
        X_c & X_d
    \end{pmatrix} \begin{pmatrix}
        \cov_\Sys & 0\\
        0 & \cov_\Mem
    \end{pmatrix}\begin{pmatrix}
        X_a^\top & X_c^\top \\
        X_b^\top & X_d^\top
    \end{pmatrix}  +  \begin{pmatrix}
        Y_a & Y_b \\
        Y_c & Y_d
    \end{pmatrix},
\end{align}
with the initial memory covariance matrix \(\cov_\Mem\).
Tracing over the memory \(\Mem\), we can read off the matrices in Eq.~\eqref{eq:map1}:
\begin{align}
    W_t = X_a, && V_t = X_b \cov_\Mem X_b^\top + Y_a,
\end{align}
and similarly we find \(\vec \lambda_t = X_b \vec l_\Mem\).
For an initial thermal memory state \(\rho_\Mem\) with \(\cov_\Mem = (\bar{n} + 1/2)\id\) and \(\vec l_\Mem = 0\), we explicitly get
\begin{align}
\label{eq:gaussian-dynamics}
    W_t = c_t \id, && V_t = [v_t +4 (\bar{n} +1/2) g^2 s_t^2] \id, && \vec \lambda_t = 0,
\end{align}
with \(c_t = \exp[-t \gamma/2] \cos\left( g t\right)\), \(s_t = \exp[-t \gamma/2]\sin(gt)/(2g)\), and \(v_t = (1-e^{-\gamma t})(\bar{N}+{1}/{2})\).

Let us consider two snapshots of this time evolution, i.e.\ a dynamics \(\Dyn = (\cpt_1, \cpt_2)\), at times \(t_{1} = \pi/(2g)\) and \(t_2 = \pi/g\).
It can intuitively be understood that these local maps on \(\Sys\) can prove the necessity of quantum memory for the behavior of the probe if we consider the idealized scenario of undamped oscillators at zero temperature (\(\gamma = \Bar{N} = \bar{n} = 0\)).
In this case, the evolution in Eq.~\eqref{eq:gaussian-dynamics} effectively describes a total deterministic loss of information at \(t_1\) and a full recovery of the probe's initial state at time \(t_2\), as illustrated in Fig.~\ref{fig:idealized-case}.
By virtue of the no-cloning theorem, such a behavior is possible only if the probe's initial state can be stored in a quantum system at the intermediate time \(t_1\).
Accordingly, this dynamics \(\Dyn = (\cpt_1, \cpt_2)\) is not realizable with classical memory and cannot be written in the form of Eq.~\eqref{eq:classical-memory}.

In a realistic scenario this idealized reasoning no longer holds and we will have to replace the heuristic argument by a proper witness of quantum memory, which is valid for cases where the decay and subsequent revival in the probe is not perfect. 
However, what is more, in an experiment, the total duration of the protocol, \(t_2 \sim \pi/g\), may be prohibitively long for the weak gravitational coupling \(g\). Thus, we need a refined protocol where the revival happens on shorter time scales. 
This can be achieved with a single local unitary intervention on the probe \(\Sys\) as shown in the following.

\subsection{The protocol}
\label{sec:protocol}
A revival in the probe system can be induced through a \emph{local} phase flip on the probe. 
We split the total protocol duration into two equal parts of length \(\tau\). At time \(t_1=\tau\), we apply a \(\pi\) rotation, \(\Pi_\Sys = \exp[i \pi a^\dagger a]\), on the probe's phase space. We treat this operation as instantaneous on the time scales
of the protocol and the gravitational interaction.
The required phase flip could in principle be implemented through a cyclic modulation of the probe’s trap frequency, using either adiabatic phase accumulation~\cite{lakeProcessingLightOptically2021} or suitably timed or shaped nonadiabatic pulses that cancel residual squeezing~\cite{skrabulisShottoshotNoiseCancellation2026,chenFastOptimalFrictionless2010}.

In the phase space coordinates it is given by a symplectic transformation \(S = -\sigma_z \otimes \id\).
Thus, the joint covariance matrix of \(\Sys\) and \(\Mem\) at the final time \(t_2 = 2\tau\) is given by
\begin{align}
    \cov_{\Sys\Mem}(t_2) = \Xcal \cov_{\Sys\Mem}(0) \Xcal^\top + \Ycal,
\end{align}
with 
\begin{align}
    \Xcal = X_{\tau}\, S\, X_{\tau}, && \Ycal = X_{\tau} \, S \, Y_{\tau} \, S^\top \, X^\top_{\tau} + Y_{\tau},
\end{align}
and \(X_\tau\) and \(Y_\tau\) as given in Eq.~\eqref{eq:X-and-Y} with \(t=\tau\).

\begin{figure}
\centering
\begin{tikzpicture}
    \node[anchor=south west, inner sep=0] (img)
        {\includegraphics[width=.9\linewidth]{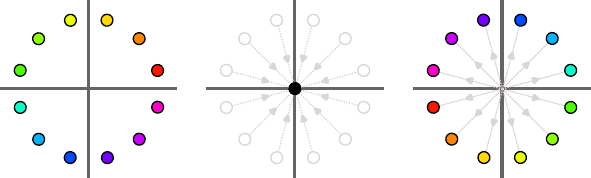}};
    \begin{scope}[x={(img.south east)}, y={(img.north west)}]
        \node at (0.22,.89) {$\scriptscriptstyle{\ket{\alpha}}$};
        \node at (0.27,.78) {$\scriptscriptstyle{\ket{\beta}}$};
        \node at (0.305,.6) {$\scriptscriptstyle{\ket{\cdots}}$};

        \node at (.782,.11) {$\scriptscriptstyle{\ket{\alpha}}$};
        \node at (0.73,.21) {$\scriptscriptstyle{\ket{\beta}}$};
        \node at (0.695,.39) {$\scriptscriptstyle{\ket{\cdots}}$};

        \node at (0.52,0.57) {\tiny{$\ket{0}$}};
    \end{scope}
\end{tikzpicture}
	\caption{Phase space visualization of the time evolution of different initial states of the probe system \(\Sys\) shown at times \(t_0 = 0\), \(t_1 = \pi/(2g)\), and \(t_2 = \pi/g\). Each circle represents a possible initial coherent state. For 
    zero damping and temperature (\(\gamma = \bar{n} = \bar{N}= 0\)),
    the maps in Eqns.~\eqref{eq:map1} and \eqref{eq:gaussian-dynamics} are given by \(\cpt_1:  W_1 = 0,
    V_1 = \id/2\), and \(\cpt_2: W_2 = -\id,  V_2 = 0\). 
    Each initial state is deterministically overwritten by the pure ground state \(\ket{0}\) given by \(\cov = \id/2\), \(\vec l  = (0,0)^\top\). Thus, no information about the initial state remains in \(\Sys\) at time \(t_1\).
    At time \(t_2\), all initial states are recovered (up to a unitary rotation in phase space).  
    Accordingly, information must have been stored in the quantum memory \(\Mem\) at intermediate times since---by virtue of the no-cloning theorem---it is impossible to create classical copies of arbitrary non-orthogonal input states. The unitary transformation of both oscillators at time \(t_1\) is actually exactly the SWAP operation (up to a phase rotation) which is the motivation for the two-sided non-LOCC witness in Ref.~\cite{lamiTestingQuantumnessGravity2024a}. Crucially, our reasoning given above does not require a test of this global dynamics but is solely based on what is observable in the probe \(\Sys\) alone.
    }
    \label{fig:idealized-case}
\end{figure}

Assuming a thermal initial memory state again,
the local probe channel \(\cpt_1\) at time \(t_1 = \tau\), right before applying the phase flip \(S\) is given by Eq.~\eqref{eq:gaussian-dynamics} as before.
For the channel \(\cpt_2\) at time \(t_2 = 2\tau\) we find
\begin{align}
\label{eq:WV-2}
    W_2 = - e^{-\gamma \tau} \id, && V_2 = v_{2\tau} \id, 
\end{align}
with \(v_{2\tau}\) as in Eq.~\eqref{eq:gaussian-dynamics}.
For the idealized zero-temperature case without damping, this is the identity map up to a local rotation and therefore identical to the second map in the example in Fig.~\ref{fig:idealized-case}.
For the case with noise we additionally get a local thermal damping. 
Importantly, the form of this map is independent of the total protocol time \(t_2 = 2\tau\) as long as the phase flip \(S\) on \(\Sys\) is applied at \(t_1 = \tau\). 
Thus, the experimenter can choose the duration of the experiment and does not have to wait until a natural revival happens at a time equivalent to the inverse coupling strength, or to detune the oscillators to get earlier revivals.

\subsection{The witness}

The necessity of quantum memory for the probe dynamics and, thus, the quantumness of the gravitational interaction can be witnessed based on measurements of first and second moments of the local channel on the probe.
Let us assume that the dynamics of the probe \(\Sys\) under interaction with the memory \(\Mem\) is given by CPT maps \(\cpt_i\). Since we do not know if the dynamics is actually of the Gaussian form derived above, we make no assumptions about the Gaussianity of these maps \(\cpt_i\).
We introduce a fictitious ancilla system \(\Anc\) and define the states
\begin{align}
\label{eq:probe-ancilla-chi}
    \chi_i = (\cpt_i \otimes \id_\Anc)[\kb{\Psi_\zeta}{\Psi_\zeta}], 
\end{align}
with the two-mode squeezed vacuum
\begin{align}
\label{eq:TMSV}
    \ket{\Psi_\zeta} = \sqrt{1-\kappa^2} \sum_{n=0}^\infty \kappa^n \ket{n}_\Sys \ket{n}_\Anc, \quad \kappa = \tanh{(\zeta/2)}.
\end{align}
We denote the covariance matrices of these states \(\chi_i\) by
\begin{align}\label{eq:joint-covariance}
    \boldsymbol\Sigma_{i} = \begin{pmatrix}
        \boldsymbol\alpha_{i} & \boldsymbol\gamma_i\\
        \boldsymbol\gamma_i^\top & \boldsymbol\beta_i
    \end{pmatrix}, \quad i = 1,2.
\end{align}
Crucially, even if the true channels \(\cpt_i\) are non-Gaussian, these covariances can be determined by measurements of first and second moments of the local time evolution on the probe \(\Sys\). 
The full states \(\chi_i\) remain unknown, our witness is solely based on the matrices \(\Sigmacov_i\).
The procedure for obtaining them is based on a weighted averaging over coherent state inputs and described in App.~\ref{app:covariance-tomography}.

The following quantity then defines a quantum memory witness:
\begin{align}
\label{eq:rigolin-witness}
    w = \Delta_2 - \Theta_1,
\end{align}
given in terms of the determinants \(\Asf_i = \det \alphacov_i\), \(\Bsf_i = \det \betacov_i\), \(\Csf_i = \det \gammacov_i\), and \(\Dsf_i = \det \Sigmacov_i\)
by 
\begin{align}
    \Theta_1 = 2 \sqrt{\Asf_1} - \sqrt{4 \Asf_1 - 1},
\end{align}
and
\begin{align}
\label{eq:Delta-2}
    \Delta_2 = \sqrt{(2\bar a_2 - P_2^+ - P_2^-)(2\bar a_2 - P_2^+ + P_2^-)},
\end{align}
with
\begin{align}
    P_2^\pm = \sqrt{\frac{\left(\sqrt{\Asf_2 \Bsf_2} \pm \abs{\Csf_2}\right)^2 - \Dsf_2}{\sqrt{\Asf_2 \Bsf_2}}}, && \bar a_2 = \frac{\sqrt{\Asf_2} + \sqrt{\Bsf_2}}{2}.
\end{align}
The probe dynamics requires
quantum memory and the gravitational interaction is quantum
if we find a \(\zeta>0\) for the squeezed state in  Eq.~\eqref{eq:TMSV} such that \(w < 0\).
The witness is inspired by the entropic one given in Ref.~\cite{backerEntropicWitnessQuantum2025a}. The derivation can be found in App.~\ref{app:rigolin-witness}.

\subsection{Scaling of the witness}
\label{sec:scaling}

For the Gaussian dynamics described by matrices \(W\) and \(V\) in Sec.~\ref{sec:protocol} the covariance matrices read
\begin{align}\label{eq:joint-covariance}
    \boldsymbol\Sigma_{i} = \begin{pmatrix}
        \boldsymbol\alpha_{i} & \boldsymbol\gamma_i\\
        \boldsymbol\gamma_i^\top & \boldsymbol\beta_i
    \end{pmatrix} = \begin{pmatrix}
        W_i A_\zeta W_i^\top+ V_i & W_i B_\zeta \\
        B^\top_\zeta W_i^\top & A_\zeta
    \end{pmatrix},\; i = 1,2,
\end{align}
with \(A_\zeta = \cosh({\zeta}) \id/2\), \(B_\zeta = \sinh({\zeta})\sigma_z/2 \). 
The value of the witness \(w\) in Eq.~\eqref{eq:rigolin-witness} for given parameters \(g, \tau, \gamma, \bar N, \bar n\) depends on the choice of \(\zeta\) in the fictitious two-mode squeezed state \(\ket{\Psi_\zeta}\) in Eq.~\eqref{eq:TMSV}, and the optimal \(\zeta\), which minimizes \(w\), has to be found numerically in general. 
We expect to see the quantumness if the damping \(\gamma\) is sufficiently weak with respect to the gravitational interaction strength \(g\).
Let us consider the zero-temperature case, \(\bar N = \bar n = 0\), first. 
In this case all expressions simplify considerably and one can show that quantum memory can be witnessed if the damping is below (see App.~\ref{app:damping-rates} for details)
\begin{align}
\label{eq:critical-gamma0}
    \gamma_0 < - \frac{2 \ln \abs{\cos g\tau}}{\tau}.
\end{align}
This expression nicely shows that in the idealized case  \(t_1 = \tau = \pi/(2g)\), i.e.\ \(\cpt_1[\rho] = \kb{0}{0}\) as in Fig.~\ref{fig:idealized-case}, vacuum damping of any strength does not destroy the quantum memory signature. 

In order to get an impression of how the magnitude of the witness grows over time, we consider the regime where the total protocol length \(t_2 = 2\tau\) is short with respect to the inverse coupling strength, i.e.\ \(g \tau \ll 1\). The optimal \(w\) can be found analytically for the zero temperature case.
We make an ansatz for the scaling of the critical damping rate \(\gamma_0\),
\begin{align}
    \gamma_0 = u g^2 \tau, \quad u > 0.
\end{align}
Expanding \(w\) in Eq.~\eqref{eq:rigolin-witness} to second order in \(\tau\) yields
\begin{align}
    w = \frac{(e^{\zeta}-1)(u e^{\zeta} -1)}{(e^{\zeta}+1)e^{\zeta}} (g\tau)^2.
\end{align}
Minimizing over \(\zeta\) gives the optimal value 
\begin{align}
    w^* =  \left(2\sqrt{2(1+u)} - u -3\right) (g\tau)^2, \quad u \leq 1.
\end{align}
The witness is negative for \(u<1\). Thus, at short times the critical noise is
\begin{align}
    \gamma_0 < g^2 \tau .
\end{align}
Not surprisingly, a longer protocol time \(t_2 = 2\tau\) increases the noise robustness and the magnitude of the witness. 

For the thermal case, \(\bar N \gg 1, \bar n > 0\), we can derive a similar relation.
This time we make the ansatz
\begin{align}
    \gamma_{\bar N} = u g^2 \tau /\bar N. 
\end{align}
The witness shows the same quadratic scaling in \(g\tau\) as before and is explicitly given by (see App.~\ref{app:damping-rates}) 
\begin{align}
\label{eq:witness-thermal}
    w = \coth\zeta
\left[
2u+
e^{-\zeta}\operatorname{sech}\zeta
\left(1+2\bar n-\cosh\zeta\right)
\right](g\tau)^2.
\end{align}
The optimal \(\zeta\) can be found numerically. 
Demanding \(w<0\) in Eq.~\eqref{eq:witness-thermal} we can read off the critical \(u\) which leads to the critical  damping rate for the thermal case
\begin{align}
\label{eq:gamma-N}
    \gamma_{\bar N} < \frac{1}{2}e^{-\zeta}\operatorname{sech}\zeta
\left(\cosh\zeta-1-2\bar n\right)\frac{g^2 \tau}{\bar N},
\end{align}
whose maximum can again be found numerically by optimizing over \(\zeta\).
In the limit \(\bar n \to 0\) it is approximately given by
\(\gamma_{\bar N} < 0.07 g^2 \tau /\bar N\).
As expected, the tolerable damping rate scales inversely with the bath temperature \(\bar N\) and quadratically in the coupling constant \(g\).
Importantly, longer total protocol times \(t_2=2\tau\) allow for stronger damping also in the thermal case.

\subsection{Asymmetric setups} 
The one-sided approach presented here is particularly valuable in asymmetric setups. While the probe \(\Sys\) requires full quantum control in order to measure its dynamics, the memory \(\Mem\) only needs to be prepared in a fixed initial state (e.g.\ by ground-state cooling in the oscillator case), a task that is significantly easier than preparing large spatial superpositions, or measuring multiple observables as commonly needed for two-sided signatures.
In order to analyze the advantage of the quantum memory approach, let us assume an experimental platform that achieves highly accurate quantum control for masses up to size \(m\). 
To measure a two-sided witness, both masses are limited to \(m\) in this case. 
In the one-sided scenario only the probe \(\Sys\) has to be of mass \(m\), while the memory mass \(M\) can be increased in order to maximize the interaction strength.

Let us define the mass ratio \(\mu = \mM/\mP\) and assume that avoiding electromagnetic interactions requires a distance \(D\) between the surfaces of the two spheres~[see Fig.~\ref{fig:setups}\,(a)]. 
We can then rewrite the coupling constant \(g\) in Eq.~\eqref{eq:osci-Hamiltonian} as
\begin{align}
\label{eq:geometric-factor}
    g = \frac{4 \pi G \varrho}{3 \omegaP}\Lambda, && \Lambda = \frac{\sqrt{\mu}}{(1 + \mu^{1/3}+D/r)^3},
\end{align}
where \(\varrho\) is the density of the masses and \(r\) is the radius of the probe.
Importantly, for given density \(\varrho\) and frequency \(\omegaP\), the coupling strength \(g\) does depend only on the mass ratio \(\mu\) and the distance ratio \(D/r\), but not on the absolute value of the masses (see also Ref.~\cite{miao_quantum_2020}). 
In scenarios where the distance \(D\) between surfaces of the oscillator masses is negligible with respect to the radius \(r\) of the probe sphere (\(D/r \approx 0)\), the geometric factor \(\Lambda\) is maximized for the symmetric case \(\mu=1\), i.e., for equal masses. This scenario is usually discussed in the limit of heavy oscillators, e.g.\ in the gram or even kilogram range~\cite{miao_quantum_2020,lamiTestingQuantumnessGravity2024a,toccaceloBenchmarksQuantumCommunication2025a}. 
However, for small masses \(m\) in the microgram range and below, the necessary surface distance \(D\) to prevent electromagnetic forces can be significantly larger than the radius \(r\) of the probe mass. In such a case, the asymmetric setup (\(\mu > 1\)) where the second mass \(M\) is substantially larger is advantageous since the maximum of the coupling strength in Eq.~\eqref{eq:geometric-factor} is  achieved for the mass ratio \(\mu = (1 + D/r)^3\).

\subsection{Experimental considerations}

Before we estimate some concrete numbers, a few general remarks are in order. In full analogy to any two-sided witness of non-LOCC gravity, such as entanglement-based approaches, the experimental setup must ensure that electromagnetic interactions between the probe and the memory are negligible compared to their gravitational coupling.
This requirement is fully analogous to that of any two-sided witness, such as in Ref.~\cite{bose_spin_2017}.
In our one-sided approach, we must additionally ensure that the observed quantum memory effects are not induced by the local environment of the probe system. Markovian damping, as considered above, is unproblematic, but electromagnetic interactions with the trap could, in principle, mimic the presence of a memory system. In practice, such parasitic effects can be excluded by comparing the witness with and without the gravitationally coupled memory system, by changing the initial state of the memory, or by varying the timing of the probe measurements.

Let us consider an exemplary case that illustrates the asymmetry of the one-sided approach and outlines where experimental advantages may emerge. As with two-sided proposals in the existing literature, current experimental technologies remain orders of magnitude removed from the required specifications. Determining the ideal experimentally viable parameter regime is beyond the scope of this paper and is an ongoing research effort that will be shaped by  technological developments.

\begin{figure}
    \centering
    \includegraphics[width=\linewidth]{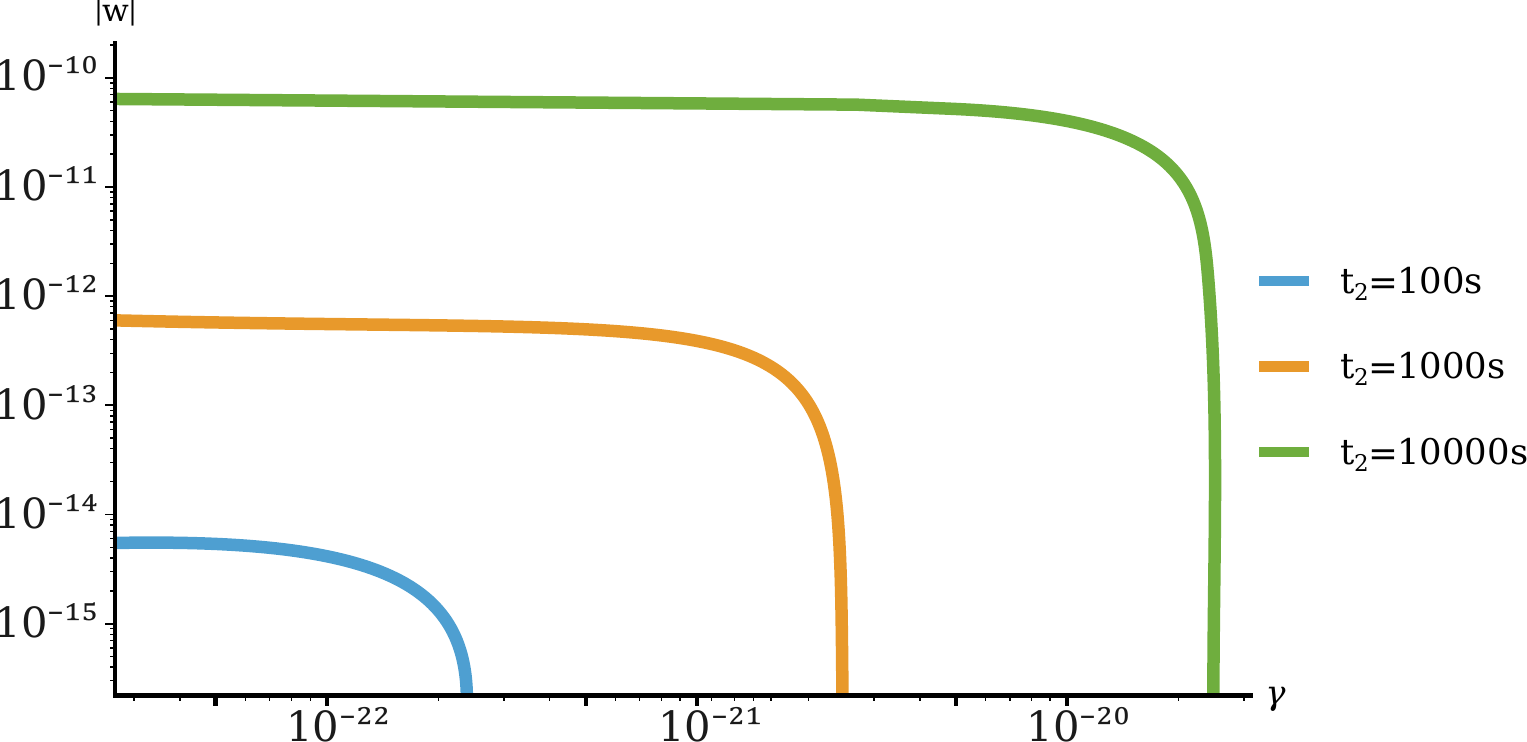}
    \caption{Magnitude of the quantum memory witness \(w\)  in Eq.~\eqref{eq:witness-thermal} as a function of the damping rate \(\gamma\) for different total protocol durations \(t_2 = 2\tau\). All other parameters as in the example in the main text.
    For each point, \(w\) is minimized over the squeezing parameter \(\zeta\) in Eq.~\eqref{eq:TMSV}. 
    Note that we plot the ranges where \(w < 0\), i.e\ where quantum memory is witnessed.
    The curve for \(t_2=1000\,\)s (orange) refers to the example in the main text. Its magnitude is at the order \(\abs{w} \sim 10^{-12}\) for most of the range \(0 < \gamma < \gamma_{\bar N}\).
    As expected from Eq.~\eqref{eq:witness-thermal}, the magnitude scales approximately quadratically in the total protocol time \(t_2 = 2\tau\). Increasing the gravitational coupling strength \(g\) by changing other experimental parameters would show the same scaling. 
    The critical damping \(\gamma_{\bar{N}}\) scales linearly with \(t_2\) in agreement with Eq.~\eqref{eq:gamma-N}.
    }
    \label{fig:w-over-gamma}
\end{figure}

We take the probe mass to be an \(m=1\)\,ng osmium sphere and we assume a distance of \(D=100\,\mu\)m between the surfaces of the oscillators, as is typical to suppress Casimir-Polder forces~\cite{bose_spin_2017}.
The optimal mass of the memory system \(\Mem\) is then \(M=0.1\)\,mg,  i.e.\ 5 orders of magnitude bigger than the probe mass. This larger mass does not need to be prepared in superposed quantum states or read out.
Let us furthermore assume oscillator frequencies of \(\omega  = 2\pi \times 0.1\,\)Hz, which leads to \(g=3.93\times 10^{-9}\)\,Hz, and a microkelvin environment, corresponding to an effective bath excitation number \(\Bar{N} = 2\times 10^{5}\). For an initial excitation \(\Bar{n} = 0.01\) in the memory and a total protocol time of \(t_2 = 1000\,\)s, the damping must be kept below \(\gamma_{\bar N} \sim 10^{-21}\)\,Hz in order to get \(w<0\) and witness the quantumness of the gravitational dynamics. While these are stringent cooling and damping requirements, they replace the need to prepare macroscopic quantum superpositions and measure an entanglement witness, which may be much more challenging  at these large mass scales.
In order to give an impression of the magnitude of the witness \(w\) in Eq.~\eqref{eq:witness-thermal}, we plot \(\abs{w}\) in Fig.~\ref{fig:w-over-gamma} as a function of the noise parameter \(\gamma\) in the range \(0<\gamma<\gamma_{\bar N}\) where the damping is below the critical value and the witness \(w\) is negative. 
For our example the witness is of order \(w \sim -10^{-12}\).

As for all non-LOCC witnesses, the strength of the signature depends significantly on the gravitational coupling strength \(g\).
Optimizing the experimental parameters such as masses, frequencies, and geometries for a given platform will be crucial for the design of a future experiment.
The quantum memory witness described here is a versatile tool which can certify the quantum nature of a dynamical channel in a wide range of systems and scenarios. 
Due to its one-sided nature, it opens up a whole new class of asymmetric scenarios that would by construction be disadvantageous for any two-sided witness.
The limiting challenge to observing the quantum dynamics of the gravitational interaction remains the very weak gravitational coupling, as it needs to overcome the environmental noise. For example, the required ultralow dissipation in the two-oscillator case is still far beyond the current state-of-the-art. However, the rapid improvements in achieving ultrahigh mechanical quality factors across diverse platforms~\cite{engelsen2024ultrahigh,de2017ultra,beccari2022strained,cupertino2024centimeter,dania2024ultrahigh,goryachev2014extremely} make such an experiment conceivable in the coming decades.
We emphasize that no other requirements on quantum state preparation, such as massive superpositions, and readout of the macroscopic system are necessary, which is the main challenge in realizing an entanglement witness \cite{bose_spin_2017}.

\section{Qubit witness}
\label{sec:qubit}
In order to illustrate the versatile nature of our witness approach we now consider a qubit probe \(\Sys\) coupled to an oscillator \(\Mem\).
This example highlights that the framework does not only apply to asymmetric setups with different masses, but that probe and memory can be entirely different quantum systems. (In App.~\ref{app:qubit-qubit} we also present a case where both \(\Sys\) and \(\Mem\) are qubits.)
The qubit \(\Sys\) is realized by a mass \(m\) that can be brought into spatial superposition of two locations denoted by \(\ket{l}\) and \(\ket{r}\) [see Fig.~\ref{fig:setups}\,(b)], an approximation widely used in the literature~\cite{bose_spin_2017,marletto_gravitationally_2017,chevalier_witnessing_2020,hanif2024testing}.
In order to fix the notation, we identify the left and right states with the eigenstates of the \(\sigma_x\) Pauli operator, i.e., \(\sigma_x \ket{l} = \ket{l}\), \(\sigma_x \ket{r} = - \ket{r}\).
This spatial qubit \(\Sys\) couples gravitationally to the position of a harmonic oscillator \(\Mem\) of mass \(M\) and frequency \(\Omega\) leading to an interaction Hamiltonian of the form  \(H_g = \hbar g \sigma_x(a^\dagger + a)\), with
\begin{align}
    g = \frac{G m \sqrt{M} d/2}{r^3\sqrt{2\hbar \Omega}},
\end{align}
where \(d\) is the spatial superposition of the qubit branches and \(r\) is their distance from the center of the oscillator.
A similar model was discussed in Refs.~\cite{carneyUsingAtomInterferometer2021, matsumura_leggett-garg_2022}.
However, the probe dynamics that follows from this interaction is of random-unitary type and can thus always be decomposed in the classical-memory form in Eq.~\eqref{eq:classical-memory}~\cite{backerLocalDisclosureQuantum2024}. 
This shortcoming can be lifted by adding a local Hamiltonian \(H_\Sys = \hbar \omega \sigma_z/2\).
Applying the rotating-wave approximation and going into a rotating frame, the problem reduces to the well-known Jaynes-Cummings model~\cite{larsonJaynesCummingsModel2024}
\begin{align}
    H =  \hbar\Delta \sigma_z/2 + \hbar g (\sigma_+ a + \sigma_- a^\dagger ),
\end{align}
where \(\Delta = \Omega - \omega\) is the detuning between the local qubit Hamiltonian and the oscillator.
We calculate the time-continuous evolution \(\cpt_t\) of the probe \(\Sys\) for an idealized noiseless case with the memory oscillator \(\Mem\) starting in the vacuum state \(\ket{0}\). The Choi state of the dynamics is defined as
\begin{align}
\label{eq:choi-state}
    E_t = (\id \otimes \cpt_t)[\kb{\phi^+}{\phi^+}],
\end{align}
with the unnormalized Bell state \(\ket{\phi^+} = \ket{00} + \ket{11}\).
It is explicitly given by (see App.~\ref{app:time-evolution})
\begin{align}
E_t ={}&|00\rangle\langle 00|
+ |p(t)|^2\,|11\rangle\langle 11|
+ \bigl[1-|p(t)|^2\bigr]\,|10\rangle\langle 10|
\nonumber\\
&+ \left[
p(t)\,|00\rangle\langle 11| + \mathrm{h.c.}
\right],
\label{eq:choi-t}
\end{align}
where
\begin{equation}
p(t)=e^{-i\Delta t/2}
\left[
\cos\frac{\tilde{\omega}t}{2}
+ i\frac{\Delta}{\tilde{\omega}}
\sin\frac{\tilde{\omega}t}{2}
\right], 
\end{equation}
with $\tilde{\omega}=\sqrt{4g^2+\Delta^2}$.
We choose \(t_1 = \pi/\tilde{\omega}\) and \(t_2 = 2\pi/\tilde{\omega}\). In the resonant case \(\Delta = 0\), this time-continuous dynamics realizes two maps
\begin{align}
\label{eq:qubit-dynamics}
    \cpt_1[\rho] = \kb{0}{0},&& \cpt_2[\rho] = \sigma_z\rho\sigma_z.
\end{align}
Similar to the resonant two-oscillator example in Fig.~\ref{fig:idealized-case}, every arbitrary input state \(\rho\) is overwritten by the pure state \(\ket{0}\) at time \(t_1\) and later recovered (up to a fixed unitary flip) at time \(t_2\).
The phase flip employed in Sec.~\ref{sec:protocol} in order to induce earlier revivals could be used here as well in full analogy to the oscillator probe case.

\begin{figure}
	\centering
\begin{tikzpicture}[scale=.9]
    \begin{scope}[tdplot_rotated_coords]
        \node (1) at (0, -3, -1.) {};
        \node (2) at (0, 0, -1.) {};
        \node (3) at (0, 3, -1.) {};
        \draw[-{Latex[length=2mm]}] (1) to [bend right=20] (2);
        \draw[-{Latex[length=2mm]}] (1) to [bend right=20] (3);
        \node (e1) at (0, -1.5, -1.0) {$\cpt_1$};
        \node (e2) at (0, 1.2, -1.2) {$\cpt_2$};
        \node (i) at (0, -3.5, 1.5) {$t_0$};
        \node (ii) at (0, 0, 1.5) {$t_1$};
        \node (iii) at (0, 3.5, 1.5) {$t_2$};
    \end{scope}
    \begin{scope}[xshift=-3.5cm, tdplot_main_coords]
        \blochdyn{0}{0}{0}{1}{2}{25}{0}{90}
    \end{scope}
    \begin{scope}[yshift=0cm, tdplot_main_coords]
        \blochdyn{0}{0}{-1.1}{0.1}{0.2}{25}{0}{90}
    \end{scope}
    \begin{scope}[xshift=3.5cm, tdplot_main_coords]
        \blochdyn{0}{0}{0}{1}{2}{25}{0}{90}
    \end{scope}
\end{tikzpicture}
    \caption{Bloch sphere visualization of the idealized probe dynamics under the gravitational coupling to the oscillator. At time \(t_1\) every input state is mapped to the pure state, \(\cpt_1[\rho_\Sys] = \kb{0}{0}\), thereby erasing all information about the initial state \(\rho_\Sys\). At time \(t_2\), the initial state is fully recovered (up to a fixed unitary), \(\cpt_2[\rho_\Sys] = \sigma_z \rho \sigma_z\). Such a dynamics is possible only if the memory stores the initial probe state \(\rho_\Sys\) at intermediate times.}
    \label{fig:bloch-spheres}
\end{figure}

As with the oscillator probe, the quantumness of the evolution could be witnessed by performing process tomography for the two maps \(\cpt_{1,2}\) and computing various witnesses, e.g.\ an entropic one~\cite{backerEntropicWitnessQuantum2025a,backerLocalDisclosureQuantum2024,backerVerifyingQuantumMemory2026}.
However, for qubit probes we can construct quantum memory witnesses that do not require a full tomography~\cite{yu_quantum_2025}. 
Instead, we measure only a few correlators of the form \(\tr\{\sigma_j \cpt_n[\sigma_i]\}\), which can be done by injecting eigenstates of \(\sigma_i\) and measuring \(\sigma_j\) at time \(t_n\).
 A witness tailored for the dynamics in Eq.~\eqref{eq:qubit-dynamics} can be found analytically. We propose
\begin{align}
\label{eq:witness-analytical}
    w = 4 + \tr \sigma_z \cpt_1[\id] + \frac{2}{3} \left( \tr\sigma_x \cpt_2[\sigma_x] - \tr \sigma_z \cpt_2[\sigma_z] \right).
\end{align}
A qubit dynamics \(\Dyn=(\cpt_1,\cpt_2)\) requires quantum memory if we find \(w<0\). 

In order to prove that \(w\) is a valid quantum memory witness, we have to show that \(w \geq 0\) for all dynamics that obey a classical decomposition as in Eq.~\eqref{eq:classical-memory}.
Using the Choi operators \(E_{1,2}\) at time \(t_{1,2}\) in Eq.~\eqref{eq:choi-t}, the witness can then be expressed as \(w = \tr[W_1 E_1] + \tr[W_2 E_2]\), with \(W_1 = \id\otimes\id + \id\otimes \sigma_z,\) and \(W_2 = \id \otimes \id +2(\sigma_x \otimes \sigma_x -\sigma_z\otimes\sigma_z)/3\).
It was shown in Ref.~\cite{yu_quantum_2025} that the operators \(W_1\) and \(W_2\) define a valid quantum memory witness if there exist positive semidefinite operators \(Q^{ADD'B}\) and \(R^{ADD'B}\) such that
\begin{align}
\label{eq:witness-decomposition}
    W_1^{AD}\otimes \frac{\id^{D'B}}{2} + W_2^{AB} \otimes \Phi_+^{DD'} = Q^{ADD'B} + (R^{ADD'B})^{\top_{D'B}},
\end{align}
where the superscripts \(A,B,D,D'\) label four different qubit Hilbert spaces, \(\Phi_+ = \kb{\phi^+}{\phi^+}\),   and \(\top_{D'B}\) denotes the partial transpose with respect to \(D'\) and \(B\). The proof can be found in Ref.~\cite{yu_quantum_2025}, Thm.~12.
Here, we only show that such a decomposition exists for our proposed witness \(w\).
Using the convention \(\sigma_z \ket{1} = + \ket{1}\), we define \(
    R = \kb{\kappa}{\kappa},\) with \( \ket{\kappa} = (\ket{1000} + \ket{0001})/\sqrt{2}\),
which is positive semidefinite by construction. It is then straightforward to verify that the operator \(Q\) resulting from Eq.~\eqref{eq:witness-decomposition} is also positive semidefinite.

\begin{figure}
    \centering
    \includegraphics[width=\linewidth]{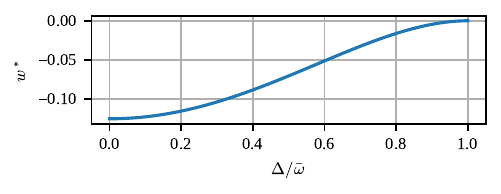}
    \caption{The numerically optimized minimal value of the witness for the form given in Eq.~\eqref{eq:SDP}. Quantum memory effects can be demonstrated for any \(\abs{\Delta}/\tilde{\omega}\). The strongest signature is obtained for the resonant case \(\Delta =0\).}
    \label{fig:numerical-witness}
\end{figure}

Despite being designed for the resonant case, the witness also works for slightly detuned dynamics in the vicinity of \(\Delta = 0\).
For larger detunings an analytical form of a suitable quantum memory witness is not known. However, we can always numerically optimize a witness that has the same structure~\cite{yu_quantum_2025}.
We are looking for Hermitian operators of the form
\begin{align}
\label{eq:numerical-form}
    W_1 &= w_{1} \id\otimes\id + w_{1z} \id \otimes \sigma_z, \\
    W_2 &= w_2 \id\otimes \id + w_{xx} \sigma_x \otimes U \sigma_x U^\dagger + w_{zz} \sigma_z\otimes\sigma_z,
\end{align}
where \( U = \exp[i \Delta\pi/(2\tilde{\omega}) \sigma_z]\) compensates a local rotation of the probe.
The \(W_{1,2}\) define the witness \(w =  \tr[W_1 E_1] + \tr[W_2 E_2]\).
Quantum memory can be demonstrated for given \(E_{1,2}\) if we find real \(w_{1},w_{1z},w_2,w_{xx},w_{zz}\) such that \(w < 0\) and the \(W_{1,2}\) form a valid witness as given by Eq.~\eqref{eq:witness-decomposition}.
This task can be solved efficiently by a semidefinite program~\cite{yu_quantum_2025}:
\begin{align}
\label{eq:SDP}
     w^* = &\min_{w_{1},w_{1z},w_2,w_{xx},w_{zz}} w \quad \text{subject to } \notag \\
    &W_1^{AD}\otimes \frac{\id^{D'B}}{2} + W_2^{AB} \otimes \Phi_+^{DD'}  \notag \\
    &\quad\quad\quad\quad = Q^{ADD'B} + (R^{ADD'B})^{\top_{D'B}},\notag \\
    &Q \succeq 0, \quad R \succeq 0, \quad \tr[W_1] + \tr[W_2] = 1.
\end{align}
The last constraint fixes a normalization for the witness. We plot the minima \(w^*\) as a function of \(\Delta/\tilde{\omega}\) in Fig.~\ref{fig:numerical-witness}, which shows that the semidefinite program yields a successful witness for any \(\Delta/\tilde{\omega}\) in the noiseless case.

Incorporating noise in the dynamics is straightforward. The numerical optimization works the same. Analogously to the oscillator probe in the previous section, the dynamics, in general, exhibits provably quantum memory effects only if the thermal damping is kept small. Besides, we note that the witness could, in principle, be further refined by incorporating multi-time correlators or even by reconstructing the full process tensor~\cite{giarmatzi_witnessing_2021,taranto_characterising_2024,yu_quantum_2025,backerVerifyingQuantumMemory2026}. However, these approaches face significantly greater experimental challenges, and we therefore do not consider them here at this stage.

\section{Conclusions}
In this paper we develop the first one-sided witness for the quantumness of gravitational dynamics. 
Similar to semi-device-independent protocols in quantum information theory, the concept only relies on the correct characterization of a probe system \(\Sys\) but makes no assumptions about the gravitationally interacting partner, here called the memory \(\Mem\).
The quantumness of the gravitational interaction is verified by measuring only the probe, showing if the dynamics exhibits quantum memory effects. This is a different capability from entanglement generation alone. 

The proposed memory witness 
extends non-LOCC tests to asymmetric setups, such as masses of different scales.  This approach can in some regimes provide tangible experimental advantages, for example when noise suppression constrains the minimal distance between the two masses.  
We also emphasize that no quantum control beyond ground state cooling and isolation is needed for the heavier partner.
Furthermore, the framework enables for the first time sufficient non-LOCC tests for scenarios where the gravitationally interacting parts are entirely different quantum systems, for example a qubit probe coupled to an oscillator memory.

Overall, our results establish a new framework in which quantum behavior can be tested using measurements on the probe system alone, thereby opening a new route to witnessing quantum features of gravity.
More broadly, the developed one-sided witness provides a general method for certifying quantum memory without Gaussianity assumptions, with potential applications in other areas of quantum information science.

\acknowledgments
We are grateful to Charlotte Bäcker for helpful discussions and support with some of the figures. We also thank Kristian Toccacelo for insightful comments on the setup. KB and IP acknowledge support by the
NSF under award No.\ 2239498, NASA under award No.\ 80NSSC25K7051 and the Sloan and Simons Foundations under award No.\ G-2023-21102. MSK acknowledges funding from the UK EPSRC through EP/Z53318X/1, EP/Y004752/1 and EP/W032643/1, the KIST
through the Open Innovation fund and the National Research Foundation of Korea grant funded by the Korean
government (MSIT) (No.\ RS-2024-00413957).

\appendix
\section{Quantum memory rules out LOCC gravity}
\label{app:LOCC}
Here we show that a global separable dynamics without quantum memory always leads to a local probe dynamics with only classical memory in the sense of Eq.~\eqref{eq:classical-memory}.
Thus, detecting quantum memory in the probe dynamics would show that gravity must be non-separable and therefore non-LOCC or that gravity itself provides a quantum memory.


A separable gravitational interaction between \(\Sys\) and \(\Mem\) is given by a Kraus representation whose operators can be decomposed into tensor products: 
\begin{align}
    \Xi[\rho] = \sum_i (A_i \otimes B_i) \rho (A_i^{\dagger} \otimes B^{\dagger}_i), 
\end{align}
with \( \sum_i (A^{\dagger}_i \otimes B_i^{\dagger})(A_i \otimes B_i) = \id. \)
We furthermore assume that the joint dynamics \((\Xi_1,\Xi_2)\) does not exploit quantum memory effects, i.e., the maps at times \(t_1\) and \(t_2\) can be written in the form
\begin{align}
\label{eq:Xi1}
    \Xi_1[\rho] &= \sum_i (A_i \otimes B_i) \rho (A_i^\dagger \otimes B_i^\dagger) \\
    \label{eq:Xi2}
    \Xi_2[\rho] &= \sum_{i,j} (C^i_j \otimes D^i_j) (A_i \otimes B_i) \rho (\ldots)^\dagger(\ldots)^\dagger,
\end{align}
with \(    \sum_j (C_j^{i\dagger} \otimes D_j^{i\dagger})(C^i_j \otimes D^i_j) = \id,\; \forall i.\)
For the probe dynamics \(\cpt_{1,2}[\rho_\Sys] = \tr_\Mem\{\Xi_{1,2}[\rho_\Sys \otimes \rho_\Mem]\}\), we can directly determine a quantum-memory-less decomposition of the form in Eq.~\eqref{eq:classical-memory}. We find
\begin{align}
    K_i = \sqrt{\tr[B_i^\dagger B_i \rho_\Mem]} A_i,
\end{align}
and the \(\Phi_i\) are given by Kraus operators
\begin{align}
G_j^i = \sqrt{\frac{{\tr[B_i^\dagger D_j^{i\dagger} D_j^i B_i \rho_\Mem]}}{{\tr[B_i^\dagger B_i \rho_\Mem]}}}C_j^i.
\end{align}

\section{Oscillator probe dynamics}\label{app:gaussian}
Here we derive the form of the local probe dynamics given by the matrices \(W_t\) and \(V_t\) in Eq.~\eqref{eq:gaussian-dynamics} in the main text.
We start from the Hamiltonian \eqref{eq:osci-Hamiltonian}, which can be rewritten as 
\(H = \hbar\vec\xi^\top \Tilde{H} \vec\xi /2\), with the vector of dimensionless quadratures for both oscillators \(\vec\xi = (q,p,Q,P)^\top\), and 
\begin{align}
    \Tilde{H} = \begin{pmatrix}
        0 & 0 & g & 0 \\
        0 & 0 & 0 & g \\
        g & 0 & 0 & 0 \\
        0 & g & 0 & 0
    \end{pmatrix} 
\end{align}
Under thermal damping with strength \(\gamma\) into a bath of average excitation \(\bar{N}\), the two-mode covariance matrix evolves according to~\cite{serafini_quantum_2023,sandulescu_open_1987}
\begin{align}\label{eq:eom}
    \dot{\cov}_{\Sys\Mem} = K 
     \cov_{\Sys\Mem} + \cov_{\Sys\Mem} K^\top + D
\end{align}
 where \(K = J \Tilde{H} - \gamma \id/2\), with \(J\) the symplectic matrix, and \(D = \gamma(\bar{N}+1/2)\id\).
The formal solution is given by~\cite{serafini_quantum_2023,sandulescu_open_1987}
\begin{align}
    \cov_{\Sys\Mem}(t) = X_t \cov_{\Sys\Mem}(0) X_t^\top + Y_t,
\end{align} with \(X_t = \exp[K t]\) and \(Y_t = \int_{0}^t X_s D X_s^\top \mathrm{d}s\), which can be computed in closed form for the case considered here.

We are interested in the reduced dynamics of the probe \(\cov = \cov_\Sys\) for a memory in an initial thermal state with \(\cov_\Mem = (\bar{n} + 1/2)\id\).
Writing \(X_t\) and \(Y_t\) in block form of \(2\times 2\) matrices  we have 
\begin{align}
    \cov_{\Sys\Mem}(t) = \begin{pmatrix}
        X_a & X_b \\
        X_c & X_d
    \end{pmatrix} \begin{pmatrix}
        \cov_\Sys & 0\\
        0 & \cov_\Mem
    \end{pmatrix}\begin{pmatrix}
        X_a^\top & X_c^\top \\
        X_b^\top & X_d^\top
    \end{pmatrix}  +  \begin{pmatrix}
        Y_a & Y_b \\
        Y_c & Y_d
    \end{pmatrix}.
\end{align}
Tracing over the memory \(\Mem\) we can read off the transformation for the probe covariance matrix \(\cov_\Sys\) given by the matrices \(W_t\) and \(V_t\) in Eq.~\eqref{eq:gaussian-dynamics}: \(W_t = X_a\) and \(V_t = X_b \cov_\Mem X_b^\top + Y_a\). The explicit result is given in the main text. 

\section{Tomography of the two-mode covariance matrix}
\label{app:covariance-tomography}
We seek the covariance matrix of the fictitious two-mode state
\begin{align}
    \chi = (\cpt \otimes \id_\Anc)[\kb{\Psi_\zeta}{\Psi_\zeta}], 
    \label{eq:coh-cov-fictitious-state}
\end{align}
with the two-mode squeezed vacuum
\begin{align}
    \ket{\Psi_\zeta} = \sqrt{1-\kappa^2} \sum_{n=0}^\infty \kappa^n \ket{n}_\Sys \ket{n}_\Anc, \quad \kappa = \tanh{(\zeta/2)}.
\end{align}
where $0<\kappa<1$ and $\mathcal E$ is the actual completely
positive, trace-preserving probe channel. No Gaussian assumption
is imposed on $\mathcal E$.

Let $\xi^\Sys=(q,p)^\top$ denote the probe quadratures
and $\xi^\Anc =(Q,P)^\top$ the ancilla quadratures.  For the quadrature vector
$\xi=(q,p,Q,P)^\top$, we define
\begin{equation}
    \Sigmacov_{jk}
    =
    \frac12
    \tr
    \left[
        \chi
        \{\Delta \xi_j,\Delta \xi_k\}
    \right],
    \qquad
    \Delta \xi_j
    =
    \xi_j-\tr[\chi \xi_j].
\end{equation}
Our objective is to reconstruct the blocks
\begin{equation}
    \Sigmacov
    =
    \begin{pmatrix}
        \alphacov & \gammacov\\
        \gammacov^\top & \betacov
    \end{pmatrix}
    \label{eq:coh-cov-target}
\end{equation}
from measurements on coherent-state inputs to $\mathcal E$.

Since the channel acts only on the probe \(\Sys\) and is trace preserving,
the ancilla marginal is unchanged. Its mean vanishes, \(\langle \xi^\Anc \rangle = 0\), and its
covariance is
\begin{equation}
    \betacov
    =
    \frac{1+\kappa^2}{2(1-\kappa^2)}\id.
    \label{eq:coh-cov-ancilla}
\end{equation}
Writing $\vec\mu
=\langle\xi^\Sys\rangle$, the remaining
blocks are determined by
\begin{align}
    \mu_j
    &=
    \tr
    \left[
        \chi(\xi^\Sys_j\otimes \id)
    \right],
    \label{eq:coh-cov-mean-target}\\
    \alphacov_{jk}
    &=
    \frac12
    \tr
    \left[
        \chi
        (\{\xi^\Sys_j,\xi^\Sys_k\}\otimes \id)
    \right]
    -\mu_j\mu_k,
    \label{eq:coh-cov-local-target}\\
    \gammacov_{jk}
    &=
    \tr
    \left[
        \chi(\xi^\Sys_j\otimes\xi^\Anc_k)
    \right].
    \label{eq:coh-cov-cross-target}
\end{align}
The block \(\gammacov\) requires no mean-subtraction term because the
ancilla mean is zero.

\subsection{Probe mean and local covariance}
To evaluate the probe-only moments in the block \(\alphacov\), we trace out the ancilla:
\begin{align}
    \tr_{\Anc}[\chi]
    = \cpt(\tr_\Anc[\kb{\Psi_\zeta}{\Psi_\zeta}])
    =  \mathcal \cpt(\tau_\nu),
\end{align}
with the local thermal state 
\begin{align}
    \tau_\nu
    =
    (1-\kappa^2)
    \sum_{n=0}^{\infty}\kappa^{2n}|n\rangle\langle n|,
    \qquad
    \nu=\frac{\kappa^2}{1-\kappa^2}.
    \label{eq:coh-cov-thermal-marginal}
\end{align}
The thermal state enters here as a tool for evaluating the
required probe moments. Its exact coherent-state decomposition is
\begin{equation}
    \tau_\nu
    =
    \int d^2\alpha\,
    P_\nu(\alpha)|\alpha\rangle\langle\alpha|,
    \qquad
    P_\nu(\alpha)
    =
    \frac{1}{\pi\nu}
    \exp\left(-\frac{|\alpha|^2}{\nu}\right).
    \label{eq:coh-cov-thermal-decomposition}
\end{equation}
By linearity,
\begin{equation}
    \mathcal E(\tau_\nu)
    =
    \int d^2\alpha\,
    P_\nu(\alpha)
    \mathcal E(|\alpha\rangle\langle\alpha|).
    \label{eq:coh-cov-output-mixture}
\end{equation}
For each coherent input state \(\ket{\alpha}\), we define the measured output mean and
symmetrized raw second moment by
\begin{align}
    r_j(\alpha)
    &=
    \tr
    \left[
        \mathcal E(|\alpha\rangle\langle\alpha|)\xi_j
    \right],
    \label{eq:coh-cov-output-mean}\\
    M_{jk}(\alpha)
    &=
    \frac12
    \tr
    \left[
        \mathcal E(|\alpha\rangle\langle\alpha|)
        \{\xi_j,\xi_k\}
    \right].
    \label{eq:coh-cov-output-second}
\end{align}
These quantities are defined for arbitrary output states,
including non-Gaussian states. Substitution into
Eqns.~\eqref{eq:coh-cov-mean-target}
and~\eqref{eq:coh-cov-local-target} yields
\begin{align}
    \vec\mu
    &=
    \int d^2\alpha\,
    P_\nu(\alpha)\vec{r}(\alpha),
    \label{eq:coh-cov-reconstructed-mean}\\
    \alphacov
    &=
    \int d^2\alpha\,
    P_\nu(\alpha)M(\alpha)
    -
   \vec\mu\vec\mu^\top.
    \label{eq:coh-cov-reconstructed-local}
\end{align}

To estimate these ensemble averages experimentally, we draw
the coherent state amplitudes \(\alpha\) independently from the
Gaussian probability density $P_\nu(\alpha)$ and prepare the states \(\ket{\alpha}\). 
The output quadrature measurements, together with the recorded
input amplitudes \(\alpha\), are then used to estimate the ensemble
averages entering the reconstructed covariance matrix.

\subsection{Probe--ancilla cross covariance}

To obtain the ancilla quadrature in
Eq.~\eqref{eq:coh-cov-cross-target}, introduce the weighted
partial trace
\begin{equation}
    K_k
    =
    \tr_{\Anc}
    \left[
        (\id\otimes\xi^\Anc_k)\kb{\Psi_\zeta}{\Psi_\zeta}
    \right].
    \label{eq:coh-cov-weighted-trace}
\end{equation}
Since $\mathcal E$ acts only on the probe, Eqns.~\eqref{eq:coh-cov-cross-target} and~\eqref{eq:coh-cov-weighted-trace} yield 
\begin{equation}
    \gammacov_{jk}
    =
    \tr
    \left[
        \xi_j^\Sys\,\mathcal E(K_k)
    \right].
    \label{eq:coh-cov-cross-reduction}
\end{equation}
It therefore remains to express $K_k$ in terms of
coherent-state projectors.

Let $a$ and $b$ be the input probe and ancilla annihilation
operators, respectively. 
Since the channel \(\cpt\) acts only on \(\Sys\) we can take the following partial traces over \(\Anc\) explicitly.
\begin{align}
    \tr_{\Anc}
    \left[
        (\id\otimes b)\kb{\Psi_\zeta}{\Psi_\zeta}
    \right] &= (1-\kappa^2)
    \sum_{n=0}^{\infty}
    \kappa^{2n+1}\sqrt{n+1}
    |n+1\rangle\langle n|
    \notag\\
    &= 
    \frac{\tau_\nu a^\dagger}{\kappa},
    \label{eq:coh-cov-b-trace}
    \end{align}
and similarly, 
\begin{align}
    \tr_{\Anc}
    \left[
        (\id\otimes b^\dagger)\kb{\Psi_\zeta}{\Psi_\zeta}
    \right] &= (1-\kappa^2)
    \sum_{n=0}^{\infty}
    \kappa^{2n+1}\sqrt{n+1}
    |n\rangle\langle n+1|\notag\\
    &=
    \frac{a\tau_\nu}{\kappa}.
    \label{eq:coh-cov-bdagger-trace}
\end{align}
Using $a|\alpha\rangle=\alpha|\alpha\rangle$ in
Eq.~\eqref{eq:coh-cov-thermal-decomposition}, we also have
\begin{align}
    a\tau_\nu
    &=
    \int d^2\alpha\,
    P_\nu(\alpha)\,
    \alpha|\alpha\rangle\langle\alpha|,\\
    \tau_\nu a^\dagger
    &=
    \int d^2\alpha\,
    P_\nu(\alpha)\,
    \alpha^*|\alpha\rangle\langle\alpha|.
\end{align}
Combining these identities with
\begin{equation}
    Q=\frac{b+b^\dagger}{\sqrt2},
    \qquad
    P=\frac{b-b^\dagger}{i\sqrt2},
\end{equation}
we obtain for the operators in Eq.~\eqref{eq:coh-cov-weighted-trace}
\begin{align}
    K_Q
    &=
    \frac{1}{\kappa}
    \int d^2\alpha\,
    P_\nu(\alpha)\,
    \sqrt2\,\operatorname{Re}\alpha\,
    |\alpha\rangle\langle\alpha|,
    \label{eq:coh-cov-KQ}\\
    K_P
    &=
    -\frac{1}{\kappa}
    \int d^2\alpha\,
    P_\nu(\alpha)\,
    \sqrt2\,\operatorname{Im}\alpha\,
    |\alpha\rangle\langle\alpha|.
    \label{eq:coh-cov-KP}
\end{align}
Applying the channel \(\cpt\) and using
Eq.~\eqref{eq:coh-cov-cross-reduction} therefore gives
\begin{align}
    \gammacov_{jQ}
    &=
    \frac{1}{\kappa}
    \int d^2\alpha\,
    P_\nu(\alpha)\,
    r_j(\alpha)\sqrt2\,\operatorname{Re}\alpha,\\
    \gammacov_{jP}
    &=
    -\frac{1}{\kappa}
    \int d^2\alpha\,
    P_\nu(\alpha)\,
    r_j(\alpha)\sqrt2\,\operatorname{Im}\alpha.
\end{align}
Thus, the fictitious probe-ancilla correlations can be obtained
by preparing coherent input states \(\ket{\alpha}\) from the thermal distribution \(P_\nu(\alpha)\), measuring their output quadratures \(r_j(\alpha)\) [see Eq.~\eqref{eq:coh-cov-output-mean}] and calculating the weighted mean above. 

In an experiment, the required ensemble averages can be
estimated from a finite number of independent runs, with
coherent input amplitudes sampled from $P_\nu$. This avoids
the need to characterize the channel on every coherent input.
Provided that the relevant ensemble-averaged fourth quadrature
moments are finite, the raw-moment estimators have
root-mean-square statistical errors scaling as $N^{-1/2}$ with \(N\) independent runs.
Together with the Gaussian sampling distribution, this
assumption also ensures that contributions from inputs with 
$|\alpha|^2\gg \nu$ to the estimated second and cross moments are exponentially suppressed.
The assumption permits non-Gaussian output states and is
satisfied by the damped harmonic-oscillator model with the
thermal states considered here.
For a concrete implementation, this assumption would need to
be supported by independent characterization of heating and
output fluctuations. A quantitative treatment of finite
sampling, preparation and detection errors, and any truncation
of the sampled amplitudes is beyond the scope of the present
proposal. An experimental certification would require these
uncertainties to be propagated through the covariance
reconstruction and the witness, with $\zeta$ fixed in advance
or chosen on independent data, and a negative witness
established at a stated confidence level.




\section{Oscillator witness}
\label{app:rigolin-witness}
Our quantum memory witness improves and simplifies the witness provided in Ref.~\cite{backerEntropicWitnessQuantum2025a} and extends its applicability to non-Gaussian dynamics.
We start from the fictitious probe-ancilla state given in Eq.~\eqref{eq:probe-ancilla-chi} 
\begin{align}
    \chi_i = (\cpt_i \otimes \id_\Anc)[\kb{\Psi_\zeta}{\Psi_\zeta}], 
\end{align}
with the two-mode squeezed vacuum
\begin{align}
    \ket{\Psi_\zeta} = \sqrt{1-\kappa^2} \sum_{n=0}^\infty \kappa^n \ket{n}_\Sys \ket{n}_\Anc, \quad \kappa = \tanh{(\zeta/2)}.
\end{align}
It was shown in Refs.~\cite{backerLocalDisclosureQuantum2024,backerEntropicWitnessQuantum2025a} that the dynamics \(\Dyn = (\cpt_1,\cpt_2)\) requires quantum memory if 
\begin{align}
\label{eq:entanglement-criterion}
    E^\sharp[\chi_1] < E[\chi_2],
\end{align}
with the entanglement of assistance~\cite{divincenzo_entanglement_1999}
\begin{align}
    E^\sharp[\chi] = \max_{\{p_i \ket{\psi_i}\}} \sum_i p_i S[\tr_\Anc \kb{\psi_i}{\psi_i}],
\end{align}
and the entanglement of formation 
\begin{align}
    E[\chi] = \min_{\{p_i \ket{\psi_i}\}} \sum_i p_i S[\tr_\Anc \kb{\psi_i}{\psi_i}],
\end{align}
where \(S\) is the von Neumann entropy and the minimization/maximization runs over all pure state decompositions, i.e.\ \(\sum_i p_i \kb{\psi_i}{\psi_i} = \chi\).
These quantities usually cannot be computed efficiently, and if the channels \(\cpt_i\) are non-Gaussian, the states \(\chi_i\) are not determined by the measured first and second moments on the probe \(\Sys\) alone. 
However, both sides of Eq.~\eqref{eq:entanglement-criterion} can be bounded by suitable quantities in order to obtain a tractable witness that is solely based on the two-mode covariance matrices of the states \(\chi_i\),
\begin{align}
\label{eq:Sigma}
    \Sigma_i = \begin{pmatrix}
        \alphacov_i && \gammacov_i \\
        \gammacov_i^\top && \betacov_i
    \end{pmatrix}.
\end{align}

We denote the Gaussian state that has the same covariance matrix as \(\chi\) by \(\chi^G\).
For the left hand side of Eq.~\eqref{eq:entanglement-criterion} we make use of the maximum-entropy principle which tells us that for given first and second moments, a Gaussian state always maximizes the entropy~\cite{holevoCapacityQuantumGaussian1999,wolfExtremalityGaussianQuantum2006},
\begin{align}
    \label{eq:maximum-entropy}
    S[\rho] \leq S[\rho^G].
\end{align}
Since the partial trace leaves local moments invariant, this establishes
\begin{align}
\label{eq:lhs-bound}
    E^\sharp[\chi_1] \leq S[\tr_\Anc \chi_1] \leq S[\tr_\Anc \chi^G_1] = S_G[\alphacov_1],
\end{align}
where the Gaussian von Neumann entropy \(S_G\) for the single-mode covariance matrix \(\alphacov\) of the probe system \(\Sys\) is defined by~\cite{serafini_symplectic_2003}
\begin{align}
\label{eq:Gaussian-entropy}
    S_G[\alphacov] = h\left(\sqrt{\det \alphacov}\right), 
\end{align}
with 
\begin{align}
    \label{eq:h-function}
    h(x) = \left(x+\frac{1}{2}\right) \ln\left(x+\frac{1}{2}\right) - \left(x-\frac{1}{2}\right)\ln\left(x-\frac{1}{2}\right).
\end{align}
This upper bound on \(E^\sharp[\chi_1]\) is used in the witness of Ref.~\cite{backerEntropicWitnessQuantum2025a} as well for the special case of Gaussian channels.

For the right-hand side of Eq.~\eqref{eq:entanglement-criterion} we employ several results from Refs.~\cite{giedkeEntanglementFormationSymmetric2003} and \cite{rigolinLowerBoundsEntanglement2004}. 
The EPR correlation of an arbitrary two-mode state with quadratures \(q,p,Q,P\) is defined by~\cite{giedkeEntanglementFormationSymmetric2003}
\begin{align}
    \Delta(\rho) = \min \left\{1, \frac{1}{2}[\Delta_{{\rho}}^2(q - Q) + \Delta_{{\rho}}^2(p + P)] \right\},
\end{align}
with the usual \(\Delta_{\rho}^2(X) = \tr[\rho X^2] - \tr[\rho X]^2 \). 
Ref.~\cite{giedkeEntanglementFormationSymmetric2003} showed that for any pure state \(\ket{\psi}\) with given EPR correlation \(\Delta(\psi)\) and the two-mode squeezed vacuum (TMSV) \(\ket{\psi^S}\) with the same \(\Delta\) we have
\begin{align}
\label{eq:entanglement-greater-squeeze}
    E[\psi] \geq E[\psi^S].
\end{align}
In other words, for a given EPR correlation \(\Delta\) the TMSV has the least entanglement among the pure states.
The entanglement for such a squeezed state is given by~\cite{giedkeEntanglementFormationSymmetric2003}
\begin{align}
    \label{eq:entanglement-squeezed-state}
    E[\psi^S] = f\left[\Delta(\psi^S)\right],
\end{align}
where
\begin{align}
\label{eq:f-function}
    f(\Delta) = y_+(\Delta) \ln y_+(\Delta) - y_-(\Delta) \ln y_-(\Delta),  
\end{align}
with
\begin{align}
\label{eq:y}
     y_\pm = \frac{1}{4}\left(\frac{1}{\sqrt{\Delta}} \pm \sqrt{\Delta}\right)^2.
\end{align}
(Note that our convention differs from Ref.~\cite{giedkeEntanglementFormationSymmetric2003} in the base of the logarithm in the function \(f\) in order to match the base in Eq.~\eqref{eq:h-function}.)
Thus, for an arbitrary pure two-mode state \(\ket\psi\) we have
\begin{align}
\label{eq:entanglement-pure-bound}
    E[\psi] \geq f[\Delta(\psi)]
\end{align}

The entanglement of formation is defined via a convex-roof construction. Thus, there is an ideal pure state decomposition \(\{(p_i, \ket{\psi_i}) \}\) such that
\begin{align}
    E[\chi_2] = \sum_i p_i E\left[\psi_i\right], && \sum_i p_i \kb{\psi_i}{\psi_i} = \chi_2.
\end{align}
With Eq.~\eqref{eq:entanglement-pure-bound} we can bound each term in this sum and find
\begin{align}
\label{eq:entanglement-to-squeeze}
    E[\chi_2] &= \sum_i p_i E[{\psi_i}] \geq \sum_i p_i f\left[\Delta({\psi_i})\right],
\end{align}
The function \(f\) is convex~\cite{giedkeEntanglementFormationSymmetric2003}, thus
\begin{align}
\label{eq:f-convex}
    \sum_i p_i f\left[\Delta({\psi_i})\right] \geq f\left[\sum_i p_i \Delta({\psi_i})\right].
\end{align}
The function \(\Delta\) is concave~\cite{rigolinLowerBoundsEntanglement2004}, thus
\begin{align}
    \sum_i p_i \Delta({\psi_i}) \leq \Delta\left[\sum_i p_i \kb{\psi_i}{\psi_i} \right] = \Delta(\chi_2).
\end{align}
Since \(f\) is decreasing in its argument~\cite{giedkeEntanglementFormationSymmetric2003}, we can combine the last three inequalities and obtain
\begin{align}
\label{eq:rhs-bound}
    E[\chi_2] \geq f\left[\Delta(\chi_2)\right].
\end{align}

We can express \(\Delta\) in terms of quantities in the covariance matrix \(\Sigmacov_2\) in Eq.~\eqref{eq:Sigma}.
The latter can be transformed by local symplectic transformations (which leave the entanglement invariant) into the standard form
\begin{align}
\label{eq:Sigma-tilde}
    \tilde\Sigmacov = \begin{pmatrix}
        \tilde\alphacov && \tilde\gammacov \\
        \tilde\gammacov^\top && \tilde\betacov
    \end{pmatrix},
\end{align}
with 
\begin{align}
\label{eq:standard-form}
    \tilde\alphacov = \begin{pmatrix}
        a && 0 \\
        0 && a
    \end{pmatrix}, && \tilde\betacov = \begin{pmatrix}
        b && 0 \\
        0 && b
    \end{pmatrix}, && \tilde\gammacov = \begin{pmatrix}
        \kappa^+ && 0 \\
        0 && \kappa^-
    \end{pmatrix},
\end{align}
where \(\abs{\kappa^+} \geq \abs{\kappa^-}\) without loss of generality.
It was shown in Ref.~\cite{rigolinLowerBoundsEntanglement2004} that EPR correlations are then given by
\begin{align}
\label{eq:Delta-derivation}
    \Delta = \min\left\{1, 2 \sqrt{(\bar a - \abs{\kappa^+})(\bar a - \abs{\kappa^-})}\right\},
\end{align}
where \(\bar a = (a+b)/2\). 
We now write \(y_\pm\) in Eq.~\eqref{eq:y} as
\begin{align}
\label{eq:nu}
      y_\pm= \nu \pm \frac{1}{2}, && \nu = \frac{1}{4}\left(\Delta + \Delta^{-1}\right),
\end{align}
and we get
\begin{align}
\label{eq:f-equals-h}
    f(\Delta) = h(\nu),
\end{align}
with \(h\) as in Eq.~\eqref{eq:h-function}.
Using the witness in Eq.~\eqref{eq:entanglement-criterion}, the two bounds in Eq.~\eqref{eq:lhs-bound} and Eq.~\eqref{eq:rhs-bound}, and Eq.~\eqref{eq:f-equals-h}, we arrive at the following witness for quantum memory in single-mode dynamics:
\begin{align}
\label{eq:h-witness}
    h(a_1) < h(\nu_2),
\end{align}
where \(a_1 = \sqrt{\det \alphacov_1}\) defined through \(\Sigmacov_1\) in Eq.~\eqref{eq:Sigma} and \(\nu_2\) defined by Eqns.~\eqref{eq:Sigma-tilde}-\eqref{eq:nu} for \(\Sigmacov_2\).

The function \(h(x)\) is increasing for \(x>1/2\). Since we have \(a_1 \geq 1/2\) (true for any valid physical state) and \(\nu_2 \geq 1/2\) (true by construction), we can replace Eq.~\eqref{eq:h-witness} by an inequality for the arguments:
\begin{align}
    \label{eq:argument-witness}
    a_1 < \nu_2.
\end{align}
Plugging in again the definition of \(\nu\) in Eq.~\eqref{eq:nu} we arrive at the witness
\begin{align}
    \Delta_2^2 - 4a_1 \Delta_2 + 1 > 0.
\end{align}
Since we have \(\Delta \leq 1\) [see Eq.\eqref{eq:Delta-derivation}], only one branch of this quadratic relation survives and we obtain the witness
\begin{align}
    \label{eq:Delta-Theta}
    \Delta_2 < \Theta_1, && \Theta_1 = 2 a_1 - \sqrt{4 a_1^2 - 1}.
\end{align}
Note that \(\Theta_1 \leq 1\). Thus, dropping the minimum function in the definition of \(\Delta\) in Eq.~\eqref{eq:Delta-derivation}, replacing it by its second argument, this witness cannot yield a false positive answer about the quantum memory effects in the dynamics. 
We finally define
\begin{align}
\label{eq:witness-appendix}
    w = \Delta_2 - \Theta_1,
\end{align}
and quantum memory is demonstrated if \(w < 0\).

Let us assume that the two covariance matrices \(\Sigmacov_{1,2}\) in Eq.~\eqref{eq:Sigma} have been determined experimentally (see App.~\ref{app:covariance-tomography}). 
In order to make the witness applicable it remains to show how the witness can be expressed explicitly in terms of the matrices \(\Sigmacov_i\).
To this end, we define 
\begin{align}
    \Asf_i = \det \alphacov_i,\; \Bsf_i = \det \betacov_i,\; \Csf_i = \det \gammacov_i, \;\Dsf_i = \det \Sigmacov_i.
\end{align}
These determinants are invariant under local symplectic transformations. 
Thus, with Eqns.~(\ref{eq:Sigma-tilde},  \ref{eq:standard-form}) we have
\begin{align}
    \sqrt{\Asf_i} &= a_i, \quad  \sqrt{\Bsf_i} = b_i, \quad  \Csf_i = \kappa_i^+\kappa_i^-.
\end{align}
Defining \(q_i = \sqrt{\Asf_i\Bsf_i}\) we get
\begin{align}
        \Dsf_i &= \left(a_i b_i - (\kappa_i^+)^2\right)\left(a_i b_i - (\kappa_i^-)^2\right)  \notag
        \\
        & = q_i^2 -q_i\left((\kappa_i^+)^2+ (\kappa_i^-)^2\right) + \Csf_i^2,
\end{align}
and thus,
\begin{align}
\label{eq:kappa-norm-sq}
    (\kappa_i^+)^2+ (\kappa_i^-)^2 = \frac{q_i^2+ \Csf_i^2 - \Dsf_i}{q_i}.
\end{align}
Since by definition \(\abs{\kappa_i^+}\geq \abs{\kappa_i^-}\), we can define two non-negative numbers
\begin{align}
\label{eq:P-raw}
    P_i^+ = \abs{\kappa_i^+} + \abs{\kappa_i^-}, && P_i^- = \abs{\kappa_i^+} - \abs{\kappa_i^-},
\end{align}
satisfying
\begin{align}
\label{eq:P-sq}
    (P_i^+)^2 &= (\kappa_i^+)^2+ (\kappa_i^-)^2 + 2\abs{\kappa_i^+\kappa_i^-},\notag\\
    (P_i^-)^2 &= (\kappa_i^+)^2+ (\kappa_i^-)^2 - 2\abs{\kappa_i^+\kappa_i^-}.
\end{align}
Since \(\abs{\kappa_i^+\kappa_i^-} = \abs{\Csf_i}\) we find with Eqns.~(\ref{eq:kappa-norm-sq}, \ref{eq:P-raw}, \ref{eq:P-sq})
\begin{align}
  P_i^\pm = \sqrt{\frac{\left(\sqrt{\Asf_i \Bsf_i} \pm \abs{\Csf_i}\right)^2 - \Dsf_i}{\sqrt{\Asf_i \Bsf_i}}}.
\end{align}
Using \(\abs{\kappa_i^\pm} =  (P_i^+ \pm P_i^-)/2\) and Eq.~\eqref{eq:Delta-derivation} we arrive at Eq.~\eqref{eq:Delta-2} in the main text.


\section{The gravitational signature}
\label{app:damping-rates}
\subsection{Zero temperature}
Here we derive the results presented in Sec.~\ref{sec:scaling}.
We start with the zero-temperature case \(\bar N = \bar n = 0\) and get for the quantities in Eq.~\eqref{eq:rigolin-witness}:
\begin{gather}
    a_1 = \frac{1}{2} + e^{-\gamma \tau} \cos^2(g \tau) \sinh^2 \frac{\zeta}{2}, \\
    \Delta_2 = \left(\cosh \frac{\zeta}{2} - e^{-\gamma \tau} \sinh \frac{\zeta}{2}\right)^2.
\end{gather}
For the resulting witness \(w\) in Eq.~\eqref{eq:rigolin-witness} we find
\begin{gather}
    \lim_{\zeta\to 0} w = 0, \\
    \lim_{\zeta\to 0} \frac{\mathrm{d}w}{\mathrm{d}\zeta} = e^{-\gamma \tau/2} \abs{\cos g \tau} - e^{-\gamma \tau}.
\end{gather}
The last equation leads to the critical damping rate in Eq.~\eqref{eq:critical-gamma0} below which we can always find a small \(\zeta\) for which the witness \(w\) becomes negative.

\subsection{Thermal case}
For the thermal case with \(\bar N \gg 1\) we make the approximation \(\bar N +1/2 \approx \bar N\) and  assume \(\gamma \ll g\).
This allows us to simplify the matrices in the generator of the dynamics  in Eq.~\eqref{eq:eom} to \(D = \gamma \bar N \id\) and \(K = J \tilde H\). 
Solving the time evolution we find for the quantities in Eq.~\eqref{eq:Delta-Theta}:
\begin{gather}
    a_1 = \bar{N}\gamma\tau
+\frac{1}{2}\cos^2(g\tau)\cosh\zeta
+\left(\frac{1}{2}+\bar{n}\right)\sin^2(g\tau), \\
\Delta_2 = e^{-\zeta} + 2 \bar N \gamma \tau,
\end{gather}
which leads to the witness \(w\) in Eq.~\eqref{eq:witness-thermal} of the main text.

\section{Time evolution of the qubit-oscillator example}

\label{app:time-evolution}

The time evolution of a pure qubit-oscillator state can be parametrized as
\begin{align}
\label{eq:joint-state}
    \ket{\Psi(t)} = \sum_n  \left(  c_{1,n}(t) \ket{1,n} + c_{0,n}(t) \ket{0,n} \right)
\end{align}
The solutions for \(c_{1,n}(t)\) and \(c_{0,n}(t)\) can be found in the literature~\cite{larsonJaynesCummingsModel2024}.
Let us assume an initial product state of qubit and oscillator
\begin{align}
    \ket{\Psi} = \ket{\psi,\phi},
\end{align}
where the oscillator state is given by 
\begin{align}
    \ket{\phi} = \sum_n c_n \ket{n}.
\end{align}
For a qubit initially in the state \(\ket{1}\) (as indicated by the superscript \((1)\) in the following equations), the time-evolved joint state \(\ket{\Psi(t)}\) in Eq.~\eqref{eq:joint-state}  is given by the coefficients
\begin{align}
\label{eq:coeff-start-1}
    &c_{1,n}^{(1)} = c_n\left[ \cos(\Omega_n t) - \frac{i\Delta}{2 \Omega_n} \sin(\Omega_n t)    \right]   e^{i \Delta t/2 } \\ \notag
    &c_{0,n}^{(1)} = - c_{n-1} \frac{i g \sqrt{n}}{\Omega_{n-1}} \sin(\Omega_{n-1} t) e^{-i \Delta t/2 }, 
\end{align}
with \(\Omega_n = \sqrt{\Delta^2/{4} + g^2(n+1)}\).
If the qubit starts in the ground state \(\ket{0}\) [indicated by superscript \((0)\)], the solution reads
\begin{align}
\label{eq:coeff-start-0}
    &c_{1,n}^{(0)}(t) = \frac{-i g \sqrt{n+1}}{\Omega_n}c_{n+1} \sin(\Omega_n t) e^{i \Delta t/2 } \\ \notag
    & c_{0,n}^{(0)}(t) = c_{n} \left[ \cos(\Omega_{n-1} t ) + \frac{ i \Delta}{2 \Omega_{n-1}} \sin(\Omega_{n-1} t)\right]  e^{-i \Delta t/2 } 
\end{align}
We are interested in the reduced dynamics of the qubit for a given initial oscillator state \(\ket{\phi} = \sum_n c_n \ket{n} \).
The map is formally given by
\begin{align}
\label{eq:map-Ecal}
    \Ecal[\rho_\Sys] = \tr_\Mem[U (\rho_\Sys \otimes \kb{\phi}{\phi})U^\dagger].
\end{align}
We compute 

\begin{align}
    E &= \tr_\Mem\left[\sum_{ijmn} \kb{i}{j} \otimes U (c_n \ket{i,n} \bra{j,m} c_m^*) U^\dagger\right]\notag\\
    &= \tr_\Mem\left[\sum_{ijmn} \kb{i}{j} \otimes \sum_{kl} c^{(i)}_{k,n} \ket{k,n} \bra{l,m} c^{(j)*}_{l,m} \right] \notag\\
    &= \sum_{ijkln} c^{(i)}_{k,n} c^{(j)*}_{l,n} \kb{ik}{jl}, 
\end{align}
where in the second line we have made use of Eqns.~(\ref{eq:coeff-start-1},\ref{eq:coeff-start-0}) to propagate the bra and ket state, and in the third line we have the partial trace over \(\Mem\) in its number basis \(\ket{n}\).
Thus, the matrix elements of the Choi state \(E\) can directly be computed from the coefficients in Eqns.~(\ref{eq:coeff-start-1},\ref{eq:coeff-start-0}).
Importantly, these coefficients depend on the initial state of the oscillator and so does the Choi state.

\section{Two-qubit example}
\label{app:qubit-qubit}

In the main text we have always considered the memory system \(\Mem\) to be an oscillator. 
Here we show a setup where both probe \(\Sys\) and memory \(\Mem\) are qubits.
We assume the two gravitationally interacting systems to be two massive particles with masses \(m\) and \(M\), respectively. 
Each particle can be brought into a spatial superposition of \(\ket{l}\), \(\ket r\) and \(\ket L\), \(\ket R\), respectively.
Thus, the quantum states of the two masses are effectively described by qubits, an approximation widely used in the literature~\cite{bose_spin_2017,marletto_gravitationally_2017,chevalier_witnessing_2020,hanif2024testing}. 
The superpositions are assumed to be horizontal and parallel with aligned centers (see Fig.~\ref{fig:qubit-qubit}).
In order to fix the notation, we identify the left and right states of each qubit with the eigenstates of the \(\sigma_x\) Pauli operator, i.e., \(\sigma_x \ket{l} = \ket{l}\), \(\sigma_x \ket{r} = - \ket{r}\), and analogously for \(\ket{L}\) and \(\ket{R}\).
The gravitational interaction between the two masses largely depends on whether the two particles are on the same side or on opposite sides leading to an effective coupling Hamiltonian
\begin{align}
\label{eq:coupling-Hamiltonian}
    H_g = -\hbar g \sigma_x\otimes\sigma_x,
\end{align}
with \(g = GMm (\xi_- - \xi_+)/(2\hbar)\), and \(\xi_\pm = [(\Delta X \pm \Delta x)^2/4 + d^2]^{-1/2}\).
This coupling \(g\) determines how quickly the two masses correlate and will therefore be the main parameter for the strength of the quantum signature. 
We see that we can basically trade a larger mass \(M\) of the memory system \(\Mem\) for a smaller mass \(m\) of the probe system \(\Sys\).

The unitary joint time evolution of the two qubits under the interaction for a duration \(\tau\) is given by the transformation
\begin{align}
\label{eq:gravitational-dynamics}
    U =\exp[-i \frac{\tau}{\hbar} H_g] = \exp[-i g {\tau}\sigma_x\otimes\sigma_x].
\end{align}
Such a unitary is entangling for almost all input states and times.
Locally, the build-up of correlations leads to a decoherence effect. 
However, the dephasing dynamics emerging from a coupling Hamiltonian like the one in Eq.~\eqref{eq:coupling-Hamiltonian} can locally always be described as a random unitary evolution~\cite{hosten_constraints_2022,maLimitsInferenceGravitational2022} which, in turn, can always be written in the form of Eq.~\eqref{eq:classical-memory} without quantum memory~\cite{backerLocalDisclosureQuantum2024}.
Thus, a dynamics that is solely generated by this gravitational interaction---as often considered in the literature---cannot be used to demonstrate quantumness by means of one-sided local measurements on the probe system only. 

\begin{figure}
    \centering
    \begin{tikzpicture}

  \draw[fill=memory, line width = .5mm] (-.75,0) circle (.4cm);
  \node at (-1.5,0) {\(\ket{L}\)};
  \draw[dashed, fill=memory,line width = .5mm] (.75,0) circle (.4cm);
  \node at (1.5,0) {\(\ket{R}\)};
  \draw[fill = darkyellow, line width = .5mm] (-1.25,-1) circle (.3cm);
  \node at (-1.8,-1) {\(\ket{l}\)};
  \draw[dashed, fill = darkyellow, line width = .5mm] (1.25,-1) circle (.3cm);
  \node at (1.9,-1) {\(\ket{r}\)};
  \draw[<->, line width = .3mm] (-.75,0) -- (.75,0) node[midway, above] {\(\Delta X\)};
  \draw[<->, line width = .3mm] (-1.25,-1) -- (1.25,-1) node[midway, below] {\(\Delta x\)};
  \draw[<->, line width = .3mm] (0,0) -- (0,-1) node[midway, right] {\(d\)};
  \node at (-2.2,0) {\(M\)};
  \node at (-2.2,-1) {\(m\)};

\end{tikzpicture}
    \caption{Two masses \(m,M\) that can be brought into spatial superposition represent two gravitationally interacting qubits.
    }
    \label{fig:qubit-qubit}
\end{figure}

The following extension of the purely gravitational (interacting) dynamics is crucial in order to obtain a local dynamics on the probe system \(\Sys\) that actually requires a quantum memory and, thus, leads to a sufficient one-sided test of quantumness.
We interrupt the joint gravitational dynamics by \emph{local} gates on the probe \(\Sys\) and the memory \(\Mem\), respectively.
These unitary transformations must \emph{not} act jointly on both parts, as this would render the whole test of the quantumness of gravity inconclusive.

The full sequence of quantum operations for this idealized model is shown in Fig.~\ref{fig:circuit}. The 
memory system starts in an equal superposition of \(\ket{L}\) and \(\ket{R}\), in our convention an eigenstate of the \(\sigma_z\) operator, i.e., \(
    \rho_\Mem = \kb{1}{1}\), with \(\ket{1} = (\ket{L}+\ket{R})/\sqrt{2}\).
Up to time \(t=\tau\), the two subsystems only interact gravitationally by the joint unitary \(U\) in Eq.~\eqref{eq:gravitational-dynamics}.
At time \(t = \tau\), local phase gates \(\phaseGate = \exp[-i \pi \sigma_z /4]\) are applied to both qubits (gates of this form can in principle be implemented by including a spin degree of freedom for each mass and applying a spin-dependent force). 
The local gates are assumed to take negligible time and the gravitational interaction \(U\) subsequently proceeds for another time interval \(\tau\).
Thus the global unitary evolution up to time \(t_1 = 2 \tau\) is given by
\begin{align}
\label{eq:Vcombined}
    V = U (\phaseGate\otimes \phaseGate) U,
\end{align}
and the reduced map for the probe system \(\Sys\) alone at this time \(t_1\) is accordingly given by \(\cpt_1[\rho_\Sys] = \tr_\Mem[V (\rho_\Sys\otimes \rho_\Mem) V^\dag ]\).
In the next step, the gate \(Z_\Sys = \sigma_z\) is applied to the probe \(\Sys\) and the gate sequence in Eq.~\eqref{eq:Vcombined} is repeated.
At time \(t_2 = 4\tau\) the local map on the probe part \(\Sys\) then reads \(
    \cpt_2[\rho_\Sys] = \tr_\Mem[V Z_\Sys V (\rho_\Sys \otimes \rho_\Mem) V^\dag Z_\Sys^\dag V^\dag]\).
The resulting dynamics \(\Dyn = (\cpt_1, \cpt_2)\) can be proven to require quantum memory as we will show below. 

\begin{figure}[t]
	\centering
\begin{quantikz}[column sep = .3cm]
\lstick{$\rho_\Sys$} \slice[label style = {yshift = -7.7em}]{$t_0=0$}
 &
 &
\gate[wires=2,style={rounded corners,fill=darkyellow}]{U} &
\gate[style={rounded corners,fill=niceblue}]{\phaseGate} &
\gate[wires=2,style={rounded corners,fill=darkyellow}]{U} 
& \slice[label style = {yshift = -7.7em}]{$t_1=2\tau$} &  
\gate[style={rounded corners, fill=nicegreen}]{Z} & 
\gate[wires=2,style={rounded corners,fill=darkyellow}]{U} & 
\gate[style={rounded corners,fill=niceblue}]{\phaseGate} & 
\gate[wires=2,style={rounded corners,fill=darkyellow}]{U}  & \slice[label style = {yshift = -7.7em}]{$t_2 = 4\tau$} & &\push{\phantom{\rho_\Sys}} \\
\lstick{$\rho_\Mem$} &
&
&
\gate[style={rounded corners,fill=niceblue}]{\phaseGate}&
&
&
&
&
\gate[style={rounded corners,fill=niceblue}]{\phaseGate}  & &  & & \push{\phantom{\rho_\Mem}}
\end{quantikz}
\begin{tikzpicture}[scale=.9]
    \begin{scope}[tdplot_rotated_coords]
        \node (1) at (0, -3, -1.) {};
        \node (2) at (0, 0, -1.) {};
        \node (3) at (0, 3, -1.) {};
        \draw[-{Latex[length=2mm]}] (1) to [bend right=20] (2);
        \draw[-{Latex[length=2mm]}] (1) to [bend right=20] (3);
        \node (e1) at (0, -1.5, -1.0) {$\cpt_1$};
        \node (e2) at (0, 1.2, -1.2) {$\cpt_2$};
    \end{scope}
    \begin{scope}[xshift=-3.5cm, tdplot_main_coords]
        \blochdyn{0}{0}{0}{1}{2}{25}{0}{90}
    \end{scope}
    \begin{scope}[yshift=0cm, tdplot_main_coords]
        \blochdyn{0}{0}{0.9}{0.2}{0.4}{25}{0}{90}
    \end{scope}
    \begin{scope}[xshift=3.5cm, tdplot_main_coords]
        \blochdyn{0}{0}{0}{1}{2}{25}{0}{90}
    \end{scope}
\end{tikzpicture}
	\caption{Circuit of the idealized example and visualization of the probe dynamics for an exemplary duration \(\tau = 0.7/g\) in Eq.~\eqref{eq:gravitational-dynamics}.  The probe \(\Sys\) and the memory \(\Mem\) interact gravitationally for a duration \(\tau\) described by the gate \(U\). Intermediate local phase gates \(S\) lead to a probe dynamics that cannot be described by classical memory as in Eq.~\eqref{eq:classical-memory}. At time \(t_1\) the probe evolution is described by an amplitude damping channel. The local \(Z\)-gate ensures that the evolution up to time \(t_1\) is essentially rewound by the subsequent gates such that the map at time \(t_2\) is the identity, i.e., \(\cpt_2[\rho_\Sys] = \rho_\Sys\). 
    }
 \label{fig:circuit}
\end{figure}

Before we do so, we take a look at the two maps \(\cpt_{1,2}\) in order to provide some intuition as to why the dynamics can actually be seen as storing and retrieving quantum information in and from the memory system \(\Mem\), similar to the two-oscillator case in the main text. 
The map at time \(t_1\) is a partial amplitude damping towards the state \(\ket{1}\) (with a rotation around the z-axis given by \(S\)):
\begin{gather}
\label{eq:damping-map}
    \cpt_1[\rho_\Sys] = \sum_{k=0}^1 S K_k \rho_\Sys K_k^\dagger S^\dagger,
\end{gather}
with \(K_0 = \sigma_+\sigma_- + {\cos(2 g \tau)} \sigma_-\sigma_+\), \(K_1 = {\sin(2g \tau) }\sigma_+\).
The map at time \(t_2\) is always the identity independently of \(g\) and \(\tau\), i.e., \(\cpt_2[\rho_\Sys] = \rho_\Sys\).

An example of the dynamics \(\Dyn=(\cpt_1,\cpt_2)\) is visualized in Fig.~\ref{fig:circuit}.
Pure input states of the probe \(\Sys\) at time \(t_0\) are mapped to a smaller ellipsoid closer to the \(\ket{1}\) state. At time \(t_2\) all states are mapped back to where they initially were at \(t_0\) and the whole Bloch sphere is recovered. The gate \(Z\) in the middle of Fig.~\ref{fig:circuit} flips the state in such a way that the evolution due to the first gate sequence \(V\) in Eq.~\eqref{eq:Vcombined} is effectively reversed by its second application.

A witness very similar to the one used in Sec.~\ref{sec:qubit} can be used to demonstrate the quantum memory effects in this dynamics,
\begin{align}
\label{eq:witness-analytical}
    w = 4 - \tr \sigma_z \cpt_1[\id] - \frac{2}{3} \left( \tr\sigma_x \cpt_2[\sigma_x] + \tr \sigma_z \cpt_2[\sigma_z] \right),
\end{align}
the only difference being sign flips. 
For our example dynamics we find
\begin{align}
\label{eq:w-value}
    w = \frac{1}{3} + \cos(4 g \tau).
\end{align}
Thus, the witness is designed to detect quantum memory for the perfect case of \(t_1 = 2\tau = \pi/(2g)\), where no information about the initial probe state remains in \(\Sys\) at \(t_1\) [full amplitude damping, see Eq.~\eqref{eq:damping-map}], and \(w\) attains its minimum. However, the same witness obviously also works in a broad range around this ideal choice of parameters.
Moreover, one can numerically show with an SDP [see Eq.~\eqref{eq:SDP}] that the same set of measured correlators can be used to construct successful witnesses for any nontrivial choice of \(\tau\) and \(g\).

\bibliography{bib}

\end{document}